\documentclass[preprint,11pt]{elsarticle}

\usepackage{geometry}
\usepackage{amssymb}
\usepackage{amsmath}
\usepackage{graphicx}
\usepackage{color}

\newcommand{\bl}[1]{{\color{blue}#1}}

\journal{Elsevier}

\begin{document}

\begin{frontmatter}

\title{Exact artificial boundary conditions of 1D semi-discretized  peridynamics}

\author[rvc]{Songsong Ji}
\author[rvy]{Gang Pang\corref{cor1}}
\author[jiwei]{Jiwei Zhang}
\author[Yibo]{Yibo Yang}
\author[Yibo]{Paris Perdikaris}
%\author[Wang]{Linjuan Wang}
%\author[rvc]{Shaoqiang Tang\corref{cor1}}
%\author[rvt]{Wing Kam Liu}%\corref{cor2}}
\address[rvy]{School of Mathematical Science, Beijing University of Areonautics and Astronautics, Beijing 100083, China. (Email: {\tt gangpang@buaa.edu.cn}).}
\address[rvc]{HEDPS,CAPT,and LTCS, College of Engineering, Peking University, Beijing 100871, China. (Email: {\tt songsong.0211@163.com}).}
\address[jiwei]{School of Mathematics and Statistics, and Hubei Key Laboratory of Computational Science, Wuhan University, Wuhan 430072, China. (Email: {\tt jiweizhang@whu.edu.cn}).}
\address[Yibo]{Department of Mechanical Engineering and Applied Mechanics, University of Pennsylvania, Philadelphia, PA, USA. (Email: {\tt pgp@seas.upenn.edu}).}
%\address[Wang]{Beijing University of Areonautics and Astronautics, Beijing, 100083, China}
%\address[rvt] {Department of Mechanical Engineering, Northwestern University, Evanston IL 60201, USA}
\cortext[cor1]{Author for correspondence, gangpang@buaa.edu.cn}
%\cortext[cof1]{Contributing equally with Yuping Ying}
%\cortext[cor2]{Author for correspondence,  w-liu@northwestern.edu}

\begin{abstract}

The peridynamic theory reformulates the equations of continuum mechanics in terms of integro-differential
equations instead of partial differential equations. It is not trivial to directly apply naive approach in artificial boundary conditions for continua to peridynamics modeling, because it usually involves semi-discretization scheme. In this paper, we present a new way to construct exact boundary conditions for semi-discretized peridynamics using kernel functions and recursive relations. Specially, kernel functions are used to characterize one single source are combined to construct the exact boundary conditions. The recursive relationships between the kernel functions are proposed, therefore the kernel functions can be computed through the ordinary differential system and integral system with high precision. The numerical results demonstrate that the boundary condition has high accuracy. The proposed method can be applied to modeling of wave
propagation of other nonlocal theories and high dimensional cases.
\end{abstract}


\end{frontmatter}

% \linenumbers

\section{Introduction}\label{}

The peridynamic theory reformulates the equations of classical continuum mechanics and have shown their application in many areas. In \cite{Sill} the peridynamics are used to deal with material
deformations, for instances, cracking, damage, and fracture problems \cite{Gerstle,Xu,Kilic,Oterkus,Oterkus2}.
In \cite{Weckner,Weckner2} wave propagation of peridynamics  in an unbounded medium are considered. In some cases, analytical methods are proposed to compute the analytical solutions for some problems \cite{Weckner,Mikata}. However, the analytical method can not to be applied for general problems such that the numerical computation is needed. In order to get a reliable numerical result of an unbounded domain in a finite computational domain, one needs to truncate the infinite
domain to a finite one and impose the boundary conditions on the boundary of computational domain. In this case, designing an artificial/absorbing boundary condition (ABC) that can well couple the numerical scheme inside the computational domain is very critical and would lead to the success of getting no reflecting numerical solutions. For instance in \cite{Weckner} the wave reflection from the artificial boundaries violate the numerical solutions. The numerical methods and analysis of boundary conditions have so far received a great attention for a long time. The similar problems come from a variety of areas in science and engineering, such as fluid mechanics, solid mechanics, acoustics,
electromagnetics, quantum mechanics, etc.
Extensive studies have been shown to ABCs in continua for every kinds of equation, such as classical heat equation, wave equation and Schr\"odinger, see
 \cite{Fevens,Xavier1, Xavier2,Anton,Greengard, Wu,Baskakov, Han1,Han2, Anton2,Li2,Besse,Besse2,Besse3,Pangs,Besse4,Engquist,Zhang,Zhang2,YP4}.
% This means one needs boundary conditions to \bl{get the solutions to} close the problem, \bl{or} Therefore, artificial/absorbing boundary conditions (ABCs) refer to those boundary conditions that are
%imposed on the artificial external boundaries to \ma{eliminate} undesired numerical reflections.

Sometimes it is not straightforward to construct the ABCs for continua models, and would involve
converting the continua model into semi-discretized version. Studies have also been devoted to the boundary conditions of molecular dynamics.
The time history treatment is a temporally global methods, and was known as the first accurate artificial boundary in molecular dynamic simulations \cite{Adelman,Cai}, which has been systematically extended to 2D and 3D lattices by Liu et al. \cite{Kadowaki,Karpov}.
The resulted bridging scale method has been widely used in many applications \cite{Park,Park1}. The time history kernel treatments are temporally global. However, it
is difficult to calculate its kernel function accurately. For exact boundary conditions of non-local PDE we refer to Zhang and Du's work \cite{Zhang,Wangjihong,Duwave}. They deal with the non-local heat, wave equation and Schr\"odinger equations through direct numerical Fourier transform. Direct  numerical Fourier transform works for decaying kernel functions such as nonlocal heat equation \cite{Zhang}. But for wave equations such as nonlocal Schr\"odinger equation, in order to preserve the accuracy heavy computation are carried on using the direct numerical Fourier transform \cite{Wangjihong}. However, for long term simulation the numerical Fourier transform is hard to be implemented due to the difficulty of controlling the numerical oscillation. In order to tackle these difficulties in developing exact boundary conditions, local ABCs have been developed, such as \cite{E, Li, Jone,Tang2, Tang3, Tang4,Tang5, Wang}. With a deep understanding of the Bessel functions, ALmost EXact (ALEX) boundary conditions
were proposed for harmonic chain \cite{Tang5}, and applied to similar systems including the Schr\"odinger
equation \cite{Tang6}. Unfortunately, the local ABC is insufficient in accuracy, it will bring apparent deviation at large time. Wang and Duan \cite{Linjuan} develop peridynamic transmitting boundary conditions (PTBCs) for 1D wave propagation by integrating the idea of local ABC \cite{Wang} into the peridynamic formulation. PTBCs have insufficient accuracy, especially in large time. It is not easy to explore the theoretical numerical precision of PTBCs due to this methods also depend on the points selection on the boundary. However PTBCs still suffers from the inaccuracy issue.

%E and Huang \cite{E,E2} proposed one such condition
%for time-discretized schemes. Li and E \cite{Li} further extended it to variational boundary conditions.
%Not long before, Wang and Tang \cite{Wang} proposed a series of simplified designs by matching the dispersion
%relation, called MBCs.

%Compared with the time history treatment, the boundary treatment of Wang
%and Tang [27] is local both in space and time.

As discussed in above, an essential task for dealing with these non-local problem is to accurately compute the kernel function accurately, because the numerical Fourier transform takes too much calculations and brings in the deviations in large time. In this paper we overcome the deficiency through the idea of the previous work \cite{Pang,Pang2}, where  the kernel functions of semi-discrete heat equation and Schr\"odinger equation on 2-D domain are computed with high accuracy. We propose the recursive
relationship between the kernel functions such that the kernel functions can be computed through the ordinary differential system and integral system rather than numerical Fourier transform. Therefore the numerical kernel functions can reach high precision. The proposed kernel function could be naturally adapted to construct the semi-discrete exact ABCs (SDEXABCs) of peridynamics and does not depend on the parameter selection. SDEXABCs can be extended into general cases for non-local semi-discrete system, and would be the corner stone in the broad area of non-local problems. For example, as is pointed out by Silling \cite{Silling4}, the governing equations of some nonlocal theories can be expressed as the form of peridynamics. SDEXABCs could also be applied to high dimensional problems of peridynamics by the same technique, one can refer to \cite{Pang,Duwave}.

The rest of the paper is organized as follows. In the next section, we construct the SDEXBC for
1D linear peridynamics. In Section 3, the validity of the algorithm for computing kernel functions is demonstrated. The applications are showed in Section 4, numerical convergence will be displayed.  Finally, conclusions are drawn in Section 5.

\section{Exact boundary conditions for  semi-discretized peridynamics}\label{sec:Central difference}
%\ma{We first introduce continue peridynamcis and its semi-discrrete systems. }

\subsection{One dimensional continue peridynamics and semi-discrete peridynamics}\label{subsec:overview}
For 1D linearized state-based peridynamics, in
the reference configuration the motion at a material point $x$ and time $t$ can be written as
\begin{equation}\label{eq:1}
  \rho \ddot{u}(x,t)=\int_{N_{x}} C(x',x)[u(x',t)-u(x,t)]dx'+b(x,t),
\end{equation}
where $u$ is the 1D longitudinal displacement and $\rho$ is the mass density. $N_{x}$ is the peridynamics neighborhood centered at point $x$ with a radius $\delta$. The horizon $\delta$ is the degree of nonlocality of the material. $x'$ represents a material point in the horizon. The $b$ is the
body force. The function $C(x',x)$ is the micromodulus function which is related to constitutive parameters of classical elasticity. For 1D structures of
isotropic elastic materials, the micromodulus functions have form of $C(x'-x)$ in the state-based and bond-based formulations \cite{Oterkus3}. Specially  in \cite{Weckner, Weckner2}, $C(x'-x)$ is given as
\begin{equation}\label{eq:2}
  C(x'-x)=\frac{4E}{\delta^3\sqrt{\pi}}e^{-(x'-x)^2/\delta^2}.
\end{equation}
The discretized form of Eq. \eqref{eq:1} is given as
\begin{equation}\label{eq:4}
\ddot{u}_{n}=\sum_{k=1}^{K}C(x_{n+k}-x_{n})   \Delta x  (u_{n+k}-u_{n}),
\end{equation}
where a uniform grid with a grid spacing $\Delta x$ is used to compute the integral in Eq. (\ref{eq:1}). Here $x_{n}$ denotes the position of a material point at $n\Delta x$, and $x_{n+k}$ denotes the position of a material point in the horizon of $x_{n}$ with $k=\pm1$, $\pm2, \cdots, \pm K$, the $\displaystyle K$ represents the nonlocal degree of the discretization.
Thus, the summation in (\ref{eq:4}) is over the whole range of $k$. Generally if $C(r)$ has little effect  on (\ref{eq:1}) when $R \geq r$, the $K=\displaystyle [R/\Delta x]$,  we can finally rewrite Eq. (\ref{eq:4})  as
\begin{equation}\label{eq:5}
\ddot{u}_{n}=a_{0}u_{n}+\sum_{k=1}^{K}a_{k}(u_{n+k}+u_{n-k})\;\bl{,}
\end{equation}
for all the integer $n$.

\subsection{Exact artificial boundary conditions for semi-discrete peridynamics}
\begin{figure}\centering
\includegraphics[width=12cm]{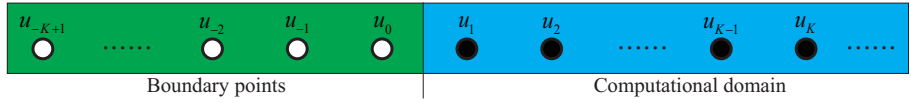}
\caption{Left boundary}\label{LFB}
\end{figure}

We now focus on  the design of ABC for (\ref{eq:5}) with the compact initial data which is confined by $1 \leq n \leq L$. The left and right boundary conditions of $u_{n}(t)$ with $-(K-1)\leq n\leq 0$ and $L+1\leq n\leq L+K$ are needed to close the ordinary differential system (\ref{eq:5}). For brevity, we only depict the left boundary in Figure  \ref{LFB} and the right boundary condition can be obtained similarly.

From Figure \ref{LFB} we need the values of $u_{n}(t)$ from $n=-(K-1)$ to $n=0$ outside the computational domain if we want to close the equation (\ref{eq:5}) on the left boundary layer. We may use the values of $u_{m}(t)$ from $m=1$ to $m=K$ in the computational domain to represent the $u_{n}(t)$, then we close (\ref{eq:5}) on the left boundary layer.

To this end, we consider the semi-infinity problem with a fixed $m$ (chosen from $1$ to $K$) as
\begin{eqnarray}\label{s2.3}
&&\ddot{f}_{n}^{m}=a_{0}f_{n}^{m}+\sum_{k=1}^{K}a_{k}\Big(f_{n+k}^{m}+f_{n-k}^{m}\Big),  \\ %\mathbb{N}\\  %\Omega_e\times(0,T]. \\
 &&f_{n}^{m}(0)=0, \dot{f}_{n}^{m}(0)=0,\quad \mbox{}\quad  n\leq 0, \\
 &&f^{m}_{k}(t)=\delta_{k}^{m}\cdot \delta(t) ,\quad \mbox{}\quad  1\leq k \leq K.
 %\label{seq2.6}&& u|_{t=0}=0,\quad \mbox{in}\quad \Omega_e=\mathbb{R}^2\backslash\Omega_{in}=\{(x,y) | 0 \leq x \leq a, 0 \leq y \leq a\},\\
%\label{seq2.7}&& u\rightarrow 0,\quad  \mbox{as}\quad |x|, |y| \rightarrow+\infty.
\end{eqnarray}
Here $\delta^{m}_{k}$ is Kronecker symbol and $\delta(t)$ is the delta function. The $f_{n}^{m}(t)$ ($n\leq 0$) is totally decided by the $K$ sources $f^{m}_{k}(t)$ with ($1 \leq m \leq K$). This is, the $f^{m}_{n}(t)$ represents the response of the $n$-th point according to the delta function at the $m$-th point.

In this way, we convolute all the $f^{m}_{n}(t)$ with a given function $F_{m}(t)$, and define $u_{n}^{m}(t)=f^{m}_{n}\ast F_{m}(t)$, we then have that the $u^{m}_{n}(t)$ satisfies
\begin{eqnarray}\label{s2.3b}
&&\ddot{u}_{n}^{m}=a_{0}u_{n}^{m}+\sum_{k=1}^{K}a_{k}\Big(u_{n+k}^{m}+u_{n-k}^{m}\Big),   \\ %\mathbb{N}\\  %\Omega_e\times(0,T]. \\
 &&u_{n}^{m}(0)=0, \dot{u}_{n}^{m}(0)=0,\quad \mbox{}\quad  n\leq 0, \\
 &&u^{m}_{k}(t)=\delta_{k}^{m}\cdot F_{m}(t) ,\quad \mbox{}\quad  1\leq k \leq K.
 %\label{seq2.6}&& u|_{t=0}=0,\quad \mbox{in}\quad \Omega_e=\mathbb{R}^2\backslash\Omega_{in}=\{(x,y) | 0 \leq x \leq a, 0 \leq y \leq a\},\\
%\label{seq2.7}&& u\rightarrow 0,\quad  \mbox{as}\quad |x|, |y| \rightarrow+\infty.
\end{eqnarray}

If we further define $u_{n}(t)=\sum_{m=1}^{K}u^{m}_{n}(t)$, we have that $u_{n}(t)$ satisfies
\begin{eqnarray}
\label{eq2:1}&&\ddot{u}_{n}=a_{0}u_{n}+\sum_{k=1}^{K}a_{k}\Big(u_{n+k}+u_{n-k}\Big),   \\
 &&u_{n}(0)=0, \dot{u}_{n}(0)=0,\quad \mbox{}\quad  n\leq 0, \\
 &&u_{m}(t)=F_{m}(t),\quad \mbox{}\quad  1\leq m \leq K.
 %\label{seq2.6}&& u|_{t=0}=0,\quad \mbox{in}\quad \Omega_e=\mathbb{R}^2\backslash\Omega_{in}=\{(x,y) | 0 \leq x \leq a, 0 \leq y \leq a\},\\
%\label{seq2.7}&& u\rightarrow 0,\quad  \mbox{as}\quad |x|, |y| \rightarrow+\infty.
\end{eqnarray}
It means that the $u_{n}(t)$ for $n\leq 0$ satisfies the governing equations of semi-discrete peridynamics with $K$ sources $F_{m}(t)$ for $1\leq m\leq K$. The $u_{n}(t)=\sum_{m=1}^{K}u^{m}_{n}(t)=\sum_{m=1}^{K}f^{m}_{n}\ast F_{m}(t)$.

In Figure \ref{LFB} the $u_{n}(t)$ ($ n \leq 0$) outside the computational domain satisfies the governing equations of semi-discrete peridynamics with $K$ sources $u_{m}(t)$ for $1\leq m\leq K$, therefore the $u_{n}(t)$ can be represented by $u_{m}(t)$ ($1\leq m \leq K$) in the computational domain, namely
\begin{eqnarray}\label{exb}
&&u_{n}(t)=\sum_{m=1}^{K}f_{n}^{m}\ast u_{m}(t) ,\quad \mbox{}\quad   -(K-1)\leq n\leq 0, 1\leq m \leq K.
\end{eqnarray}
If the $f^{m}_{n}(t)$ is known, we take the relationship \eqref{exb} as the exact ABCs on the left boundary layer.
Similarly, we also can achieve the exact ABC on the right boundary layer in the form of
\begin{eqnarray}\label{exbr}
&&u_{L-n+1}(t)=\sum_{m=1}^{K}f_{n}^{m}\ast u_{L-m+1}(t) ,\quad \mbox{}\quad   -(K-1)\leq n\leq 0, 1\leq m \leq K.
\end{eqnarray}

In the remainder, we focus on how to determine the kernel function $f^{m}_{n}(t)$. To do so,  we introduce the discrete Fourier transform $U(s)$ with a fixed $m$ as
\begin{eqnarray}\label{U}
U(s)=\displaystyle\sum_{n=-\infty }^{0}e^{inx}\tilde{f}_{n}^{m}(s),
\end{eqnarray}
where $\tilde{f}_{n}^{m}(s)$ is the Laplace transform of $f_{n}^{m}(t)$ in the time direction.

Thus we have
\begin{eqnarray*}
&&s^2U=a_{0}U+\sum_{k=1}^{K}a_{k}(\sum_{n=-\infty}^{0}e^{inx}\tilde{f}_{n+k}^{m}+\sum_{n=-\infty}^{n=0}e^{inx}\tilde{f}^{m}_{n-k})\\
&&=a_{0}U+\sum_{k=1}^{K}a_{k}(e^{-ikx}\sum_{l=-\infty}^{k}e^{ilx}\tilde{f}^{m}_{l}+e^{ikx}\sum_{l=-\infty}^{-k}e^{ilx}\tilde{f}^{m}_{l})\\
&&=(a_{0}+2\sum_{k=1}^{K}a_{k}\cos kx)U+\sum_{k=1}^{K}a_{k}(e^{-ikx}\sum_{l=1}^{k}e^{ilx}\tilde{f}_{l}^{m}-e^{ikx}\sum_{l=-(k-1)}^{0}e^{ilx}\tilde{f}_{l}^{m})\\
&&=(a_{0}+2\sum_{k=1}^{K}a_{k}\cos kx)U+\sum_{k=1}^{K}a_{k}e^{-ikx}\sum_{l=1}^{k}e^{ilx}\tilde{f}_{l}^{m}-\sum_{k=1}^{K}a_{k}e^{ikx}\sum_{l=1}^{k}e^{-i(l-1)x}\tilde{f}_{-l+1}^{m}\\
&&=(a_{0}+2\sum_{k=1}^{K}a_{k}\cos kx)U-\sum_{l=1}^{K}\Big(\sum_{k=l}^{K}a_{k}e^{i(k+1)x}e^{-ilx}\Big)\tilde{f}_{-l+1}^{m}+\sum_{l=1}^{K}\Big(\sum_{k=l}^{K}a_{k}e^{-ikx}e^{ilx}\Big)\tilde{f}^{m}_{l}\\
&&=(a_{0}+2\sum_{k=1}^{K}a_{k}\cos kx)U-\sum_{l=0}^{K-1}\Big(\sum_{k=l+1}^{K}a_{k}e^{i(k-l)x}\Big)\tilde{f}_{-l}^{m}+\sum_{l=1}^{K}\Big(\sum_{k=l}^{K}a_{k}e^{i(l-k)x}\Big)\tilde{f}^{m}_{l}\\
&&=(a_{0}+2\sum_{k=1}^{K}a_{k}\cos kx)U-\sum_{l=-(K-1)}^{0}\Big(\sum_{k=1-l}^{K}a_{k}e^{i(k+l)x}\Big)\tilde{f}_{l}^{m}+\sum_{l=1}^{K}\Big(\sum_{k=l}^{K}a_{k}e^{i(l-k)x}\Big)\tilde{f}^{m}_{l}.
\end{eqnarray*}
Noting the facts that $\tilde{f}^{m}_{l}=\delta^{m}_{l}$, we then have
\begin{eqnarray}
&&s^2U=(a_{0}+2\sum_{k=1}^{K}a_{k}\cos kx)U-\sum_{l=0}^{K-1}\Big(\sum_{k=l+1}^{K}a_{k}e^{i(k-l)x}\Big)\tilde{f}_{-l}^{m}+\sum_{l=1}^{K}\Big(\sum_{k=l}^{K}a_{k}e^{i(l-k)x}\Big)\tilde{f}^{m}_{l}\\
&&=(a_{0}+2\sum_{k=1}^{K}a_{k}\cos kx)U-\sum_{l=-(K-1)}^{0}\Big(\sum_{k=1-l}^{K}a_{k}e^{i(k+l)x}\Big)\tilde{f}_{l}^{m}+\sum_{k=0}^{K-m}a_{k+m}e^{-ikx},
\end{eqnarray}
which leads to
\begin{eqnarray*}
U(s)&=&-\sum_{l=-(K-1)}^{0}\Big(\sum\limits_{k=1-l}^{K}a_{k}\frac{e^{i(l+k)x}}{s^2-a_{0}-2\sum_{k=1}^{K}a_{k}\cos kx}\Big)\tilde{f}_{l}^{m}(s) \\
&&+\sum\limits_{k=0}^{K-m}a_{k+m}\frac{e^{-ikx}}{s^2-a_{0}-2\sum_{k=1}^{K}a_{k}\cos kx}.
\end{eqnarray*}

By introducing
$$\tilde{g}_{m}(s)=\displaystyle \frac{1}{2\pi}\displaystyle\int_{-\pi}^{\pi}\frac{e^{-imx}}{s^2-a_{0}-2\sum_{k=1}^{K}a_{k}\cos kx}dx,$$
 and from $\displaystyle \tilde{f}_{n}^{m}(s)=\frac{1}{2\pi}\int_{-\pi}^{\pi}U(s) e^{-i n x}dx$ and $\tilde{g}_{m}(s)=\tilde{g}_{-m}(s)$, we have
\begin{eqnarray}
&&\tilde{f}^{m}_{n}=-\sum_{l=-(K-1)}^{0}\Big(\sum_{k=1-l}^{K}a_{k}\tilde{g}_{k+l-n}\Big)\tilde{f}_{l}^{m}+\sum_{k=0}^{K-m}a_{k+m}\tilde{g}_{k+n}
\end{eqnarray}
with $ \mbox{}  1\leq m \leq K, -(K-1)\leq n \leq 0$, which generate a linear integral system
\begin{eqnarray}\label{f}
&&f^{m}_{n}=-\sum_{l=-(K-1)}^{0}\sum_{k=1-l}^{K}a_{k}(g_{k+l-n}\ast f_{l}^{m})+\sum_{k=0}^{K-m}a_{k+m}g_{k+n}, \\
&&\quad \mbox{}\quad  1\leq m \leq K, -(K-1)\leq n \leq 0.
\end{eqnarray}

From $g_{m}(t)=g_{-m}(t)$ we only need all the $g_{m}(t)$s with $0 \leq m \leq 2K-1$ to determine $f^{m}_{n}(t)$. It is obvious to see for any integer $m $,
\begin{eqnarray}
\label{s2.3}&&s^2\tilde{g}_{m}-\delta_{m,0}=a_{0}\tilde{g}_{m}+\sum_{k=1}^{K}a_{k}\Big(\tilde{g}_{m+k}+\tilde{g}_{m-k}\Big),\quad \mbox{}\quad  %n\geq 0   \\ %\mathbb{N}\\  %\Omega_e\times(0,T]. \\
% &&f^{m}_{k}(t)=\delta_{-m,k}(t) ,\quad \mbox{}\quad  n\geq 0, -K\leq k \leq -1
 %\label{seq2.6}&& u|_{t=0}=0,\quad \mbox{in}\quad \Omega_e=\mathbb{R}^2\backslash\Omega_{in}=\{(x,y) | 0 \leq x \leq a, 0 \leq y \leq a\},\\
%\label{seq2.7}&& u\rightarrow 0,\quad  \mbox{as}\quad |x|, |y| \rightarrow+\infty.
\end{eqnarray}
and the application of inverse Laplace transform  leads to
\begin{eqnarray}\label{geq}
\ddot{g}_{m}(t)&=&a_{0}g_{m}+\sum_{k=1}^{K}a_{k}\Big(g_{m+k}+g_{m-k}\Big),\\
\dot{g}_{m}(0)&=&\delta_{0}^{m}.  % ,\quad \mbox{}\quad  n\geq 0 %\mathbb{N}\\  %\Omega_e\times(0,T]. \\
 %\label{seq2.6}&& u|_{t=0}=0,\quad \mbox{in}\quad \Omega_e=\mathbb{R}^2\backslash\Omega_{in}=\{(x,y) | 0 \leq x \leq a, 0 \leq y \leq a\},\\
%\label{seq2.7}&& u\rightarrow 0,\quad  \mbox{as}\quad |x|, |y| \rightarrow+\infty.
\end{eqnarray}
On the other hand,
\begin{eqnarray}
&&\frac{d}{ds}\sum_{k=1}^{K}k a_{k}(\tilde{g}_{m-k}-\tilde{g}_{m+k})=\frac{d}{ds}\left(\frac{1}{2\pi}\int_{-\pi}^{\pi}\frac{2i\sum_{k=1}^{K}k a_{k}\sin kx e^{-imx} }{s^2-a_{0}-2\sum_{k=1}^{K}a_{k}\cos kx}dx\right)\nonumber\\
 &&=-\frac{2s}{2\pi}\int_{-\pi}^{\pi}\frac{2i\sum_{k=1}^{K}k a_{k}\sin kxe^{-imx} }{(s^2-a_{0}-2\sum_{k=1}^{K}k a_{k}\cos kx)^2}dx\nonumber\\
 &&=-\frac{2s}{2\pi}\int_{-\pi}^{\pi}\frac{ie^{-imx} }{(s^2-a_{0}-2\sum_{k=1}^{K}a_{k}\cos kx)^2}d(s^2-a_{0}-2\sum_{k=1}^{K}a_{k}\cos kx)\nonumber\\
 &&=-\frac{ 2ms}{2\pi}\int_{-\pi}^{\pi}\frac{e^{-imx} }{s^2-a_{0}-2\sum_{k=1}^{K}a_{k}\cos kx}dx=-2 ms\tilde{g}_{m},
\label{s2.3}
\end{eqnarray}
which means
\begin{eqnarray}
\label{s2.3}&&\sum_{k=1}^{K}k a_{k}(g_{m+k}(t)-g_{m-k}(t))=-\frac{2 m}{t}\dot{g}_{m}(t),
\end{eqnarray}
we can rewrite it as
\begin{eqnarray}\label{gb}
&&g_{m}(t)=g_{m-2K}(t)-\sum_{k=1}^{K-1}\frac{k a_{k}}{K a_{K}}(g_{m-K+k}(t)-g_{m-K-k}(t))-\frac{2(m-K)}{K a_{K} t}\dot{g}_{m-K}(t).
\end{eqnarray}
The above is the recursive relationship between $g_{m}(t)$ and $\dot{g}_{m}(t)$.

Thus $g_{m}(t)$ can be computed through (\ref{geq}) with $0 \leq m \leq 2K-1$. The left boundary $g_{m}(t)$ with $-K \leq m \leq 0$ can be computed by $g_{m}(t)=g_{-m}(t)$ and the right boundary $g_{m}(t)$ with $2K \leq m \leq 3K-1$ can be computed by recursive relationship (\ref{gb}). Therefore we determine the ordinary system of $g_{m}(t)$ with $0 \leq m \leq 2K-1$. Then we can determine $f^{m}_{n}(t)$ from integral system (\ref{f}). We point out that the above methodology can be extended into high dimensional case, see \cite{Pang} for the design of ABCs for two-dimensional Schr\"odinger equations.

\section{Numerical example of computing kernel functions}

We now use semi-discrete Euler-Bernoulli beam to illustrate the computing method of kernel functions. As one of the elemental objects in solid mechanics, the Euler-Bernoulli beam has wide applications, such as in railway and bridge engineering, etc. Describing bending wave propagation in an elastic rod, Euler-Bernoulli theory gives an equation for the transverse displacement $u(x,t)$ of
an undriven beam along its length direction $x$ as
\begin{eqnarray}\label{EBB}
&&\rho A u_{tt}+(EI u_{xx})_{xx}=0.
\end{eqnarray}
Here, $\rho$, $A$, $E$ and $I$ are the material density, the area of the beam cross section, the Young modulus, and the bending moment of inertia, respectively.
Discretizing with a uniform space grid size $\Delta x$, we rescale time by $\displaystyle\sqrt{\frac{EI}{\rho A}}\frac{1}{\Delta x^2}$
to get the non-dimensionalized semi-discrete equation for $u_{n}(t)=u(n\Delta x,t)$ as follows.
\begin{eqnarray}\label{euler}
&& \ddot{u}_{n}=-u_{n-2}+4u_{n-1}-6u_{n}+4u_{n+1}-u_{n+2}.
\end{eqnarray}

In this case, we compute an accurate numerical approximation
of $g_{n}(t)$ for $K=2$ with $a_{0}=-6$, $a_{1}=4$ and $a_{2}=-1$ which comes from the semi-discretized (\ref{EBB}). Therefore
\begin{eqnarray}
\label{s2.3}&&g_{m}(t)=g_{m-4}(t)-\frac{a_{1}}{2 a_{2}}(g_{m-1}(t)-g_{m-3}(t))-\frac{2(m-2)}{2 a_{2} t}\dot{g}_{m-2}(t).
\end{eqnarray}
By taking $m=4$ we have
\begin{eqnarray}\label{B11}
&g_{4}(t)=g_{0}(t)-\displaystyle\frac{a_{1}}{2 a_{2}}(g_{3}(t)-g_{1}(t))-\frac{2}{ a_{2} t}\dot{g}_{2}(t),
\end{eqnarray}
and
\begin{eqnarray}
g_{5}(t)&=&g_{1}(t)-\displaystyle\frac{a_{1}}{2 a_{2}}(g_{4}(t)-g_{2}(t))-\frac{3}{ a_{2} t}\dot{g}_{3}(t)\nonumber\\
&=&g_{1}(t)+\displaystyle\frac{a_{1}}{2 a_{2}}g_{2}(t)-\frac{3}{ a_{2} t}\dot{g}_{3}(t)-\frac{a_{1}}{2 a_{2}}g_{4}(t).
\label{B2}
\end{eqnarray}
We combine the closed conditions (\ref{B11}), (\ref{B2}) together with equation
\begin{eqnarray*}
 \ddot{g}_{m}(t)&=&a_{0}g_{m}+\sum_{k=1}^{2}a_{k}\Big(g_{m+k}+g_{m-k}\Big),\\
 \dot{g}_{m}(0) &=&\delta_{0}^{m}.  \quad \mbox{}\quad  0\leq n \leq 3,
 \end{eqnarray*}
and $g_{m}(t)=g_{-m}(t)$ to get the closed ordinary differential system.

We use the Runge-Kutta 4 (RK4) scheme and closed relations (\ref{B11}), (\ref{B2}) to compute $g_{m}(t)$. Let us also remark that other schemes than RK4 could be used, but of
order larger than 2. After the computation of $g_{m}(t)$, we can compute the $f^{m}_{n}(t)$ through (\ref{f}) by trapezoidal integral, reads
\begin{eqnarray}\label{scheme1}
f^{m,j}_{n}=-\sum_{l=0}^{K-1}\sum_{k=l+1}^{K}a_{k}\Delta t \left(\sum_{\alpha=0}^{j}g_{k+n-l}^{\alpha} f_{l}^{m,j-\alpha}-\frac{g_{k+n-l}^{0} f_{l}^{m,j}+g_{k+n-l}^{j} f_{l}^{m,0}}{2}\right)
+\sum_{k=0}^{K-m}a_{k+m}g_{k-n}^{j}.
\end{eqnarray}
where $f^{m,j}_{n}$ is the numerical solution of $f^{m}_{n}(j\Delta t)$ and $g^{j}_{n}(t)$ the numerical solution of $g_{n}(j\Delta t)$. According to $g_{n}(0)=0$ for any integer $n$, one has $g_{k+n-l}^{0}=0$ in (\ref{scheme1}), which leads to explicit scheme,
\begin{eqnarray}\label{scheme2}
f^{m,j}_{n}=-\sum_{l=0}^{K-1}\sum_{k=l+1}^{K}a_{k}(\sum_{\alpha=1}^{j}g_{k+n-l}^{\alpha} f_{l}^{m,j-\alpha}\Delta t-g_{k+n-l}^{j} f_{l}^{m,0}\Delta t/2)+\sum_{k=0}^{K-m}a_{k+m}g_{k-n}^{j}.
\end{eqnarray}

%which follows the numerical scheme
%\begin{eqnarray}\label{Dg}
%&&f^{m,j}_{n}=-\sum_{l=0}^{K-1}\sum_{k=l+1}^{K}a_{k}(\sum_{\alpha=0}^{j}g_{k+n-l}^{\alpha} f_{l}^{m,j-\alpha}\Delta t-g_{k+n-l}^{0} f_{l}^{m,j}\Delta t/2-g_{k+n-l}^{j} f_{l}^{m,0}\Delta t/2)\\
%&&+\sum_{k=0}^{K-m}a_{k+m}g_{k-n}^{j}.\\
%\end{eqnarray}
%Because $g_{m}(0)=0$ for any integer $n$, the scheme (\ref{Dg}) can be simplified as
%\begin{eqnarray}
%&&f^{m,j}_{n}=-\sum_{l=0}^{K-1}\sum_{k=l+1}^{K}a_{k}(\sum_{\alpha=1}^{j}g_{k+n-l}^{\alpha} f_{l}^{m,j-\alpha}\Delta t-g_{k+n-l}^{j} f_{l}^{m,0}\Delta t/2)+\sum_{k=0}^{K-m}a_{k+m}g_{k-n}^{j}\\
%\end{eqnarray}
%where $f^{m,j}_{n}$ is the numerical solution of $f^{m}_{n}(j\Delta t)$.

\begin{figure}\centering
\includegraphics[width=8cm]{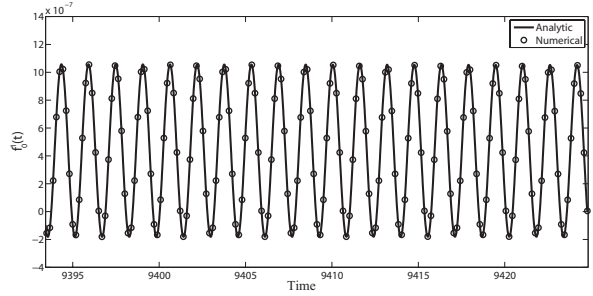}
\caption{Numerical solution and analytic solution for $f^{1}_{0}(t)$ on $[2990\pi, 3000\pi]$}\label{kcompare}
\end{figure}

To illustrate
the validity of the algorithm, we compare the numerical $f^{1}_{0}(t)$ and the analytic solution $f^{1}_{0}(t)=\displaystyle\frac{2 J_{1}(2t)}{t}\sin(2t)$  in the large time interval $ [2990\pi, 3000\pi]$, the results are displayed in Figure  \ref{kcompare}.

 The $l_{\infty}$-error
 \[\displaystyle\max_{-1\leq n\leq 0, 1\leq m\leq 2}|f^{m,analytic}_{n}-f^{m}_{n}|,\]
  is used to measure the performance. The computed convergence rates of the numerical approximation are reported
in Table \ref{converkernel}. As the time step  decreases, the approximation
quality of $f^{m}_{n}(t)$ increases. The convergent order is big than 2. The algorithm is then able to
efficiently and accurately compute $f^{m}_{n}(t)$ even for large values of
$t$. The proposed
algorithm considerably enhances the accuracy and efficiency for computing the kernel functions, avoids the numerical deviation of numerical Fourier transform in large time.

\begin{table}[h]
\begin{center}
\caption { $l_{\infty}$-error and convergence rate for different time.}\label{converkernel}
\begin{tabular}{|c|c|c|c|c|c|c}\hline
$l_{\infty}$-error &$\Delta t=3\pi/500$&$\Delta t=3\pi/400$&$\Delta t=3\pi/250$& $\Delta t=3\pi/200$&rate \\\hline
$t=3\pi$     &  9.26e-06 & 1.63e-05  & 4.79e-05 & 7.88e-04 &2.32\\\hline
$t=18\pi$    &  9.31e-07 & 1.71e-06  & 5.92e-06 & 1.11e-05 &2.69\\\hline
$t=300\pi$     &  3.80e-08 & 8.71e-08  & 5.15e-07& 1.24e-06&3.80\\\hline
$t=3000\pi$    &  8.51e-09 & 2.16e-08  & 1.52e-07 & 3.93e-07 &4.17 \\\hline
\end{tabular}
\end{center}
\end{table}

\section{Numerical Tests}\label{sec:ad_eq}
\subsection{Bar}\label{bar}
In \cite{Weckner,Weckner2}, unsuitable boundary condition violate the numerical solutions. In \cite{Linjuan}, the numerical solution deviate the analytic solution apparently at long time. Here we consider the same problem for an infinite bar. From \cite{Weckner,Weckner2,Linjuan}, the micromodulus function can be taken as
\begin{equation}\label{modu}
  C(x'-x)=\frac{4}{\delta^3\sqrt{\pi}}e^{-(x'-x)^2/\delta^2},
\end{equation}
and the initial displacement and the initial velocity can be written as:
\begin{eqnarray}
&&u(x,0)=e^{-x^2}, \quad  \dot{u}(x,0)=0.
\end{eqnarray}

The peridynamic analytical solution of this initial value problem is
\begin{equation}\label{ana}
  u(x,t)=\frac{1}{\sqrt{\pi}}\int_{0}^{\infty}\cos(kx)e^{-k^2/4}\cos\Big(t\sqrt{\frac{1-e^{-k^2\delta^2/4}}{\delta^2/4}}\Big)dk.
\end{equation}
%As $\delta \rightarrow 0$, the solution (\ref{ana}) converges to the well-known DAlembert-solution
%\begin{equation}\label{dalembert}
%  \lim_{\delta\rightarrow0}u(x,t)=\frac{1}{2}(e^{-(x-t)^2}+e^{(x+t)^2}).
%\end{equation}
 As what shown in \cite{Linjuan}, the computational domain can be chosen as a non-dimensional length 20, that is, $x\in [-10,10]$. At the two ends of the segment we use the SDEXABC. The space step can be chosen as $\Delta x=0.1$, and $\delta$ can be chosen as $0.25$. The micromodulus functions (\ref{modu}) is used for $|x'-x|\leq 0.75$ and 0 elsewhere, in this case $K=\displaystyle [0.75/\Delta x ]=7$.
The inner scheme is performed with a velocity Verlet algorithm
\begin{eqnarray}
&&u^{n+1}_{j}=u^{n}_{j}+\dot{u}^{n}_{j}\Delta t+\ddot{u}^{n}_{j}\Delta t^2/2,\\
&&\dot{u}^{n+1}_{j}=\dot{u}^{n}_{j}+\Delta t/2(\ddot{u}^{n}_{j}+\ddot{u}^{n+1}_{j}).
\end{eqnarray}
Following (\ref{exb}) and (\ref{exbr}), the boundary conditions read as
\begin{eqnarray}\label{b1}
&&u_{n}(t)=\sum_{m=1}^{K}f_{n}^{m}\ast u_{m}(t) ,\\
&&u_{L-n+1}(t)=\sum_{m=1}^{K}f_{n}^{m}\ast u_{L-m+1}(t) ,\quad \mbox{}\quad   -(K-1)\leq n\leq 0, 1\leq m \leq K.
\end{eqnarray}
We use the trapezoidal integral to compute the convolutions, namely
\begin{eqnarray}\label{convscheme}
&&u_{n}^{j}=(\sum_{m=1}^{K}\sum_{\alpha=1}^{j-1}f_{n}^{m,\alpha} u_{m}^{j-\alpha})\Delta t+ \sum_{m=1}^{K}(f_{n}^{m,0} u_{m}^{j}+f_{n}^{m,j} u_{m}^{0})\Delta t/2,\\
&&u_{L-n+1}^{j}=(\sum_{m=1}^{K}\sum_{\alpha=1}^{j-1}f_{n}^{m,\alpha} u_{L-m+1}^{j-\alpha})\Delta t+ \sum_{m=1}^{K}(f_{n}^{m,0} u_{L-m+1}^{j}+f_{n}^{m,j} u_{L-m+1}^{0})\Delta t/2, \nonumber\\
&&-(K-1)\leq n\leq 0, 1\leq m \leq K.
\end{eqnarray}

\begin{figure}\centering
\includegraphics[width=8cm]{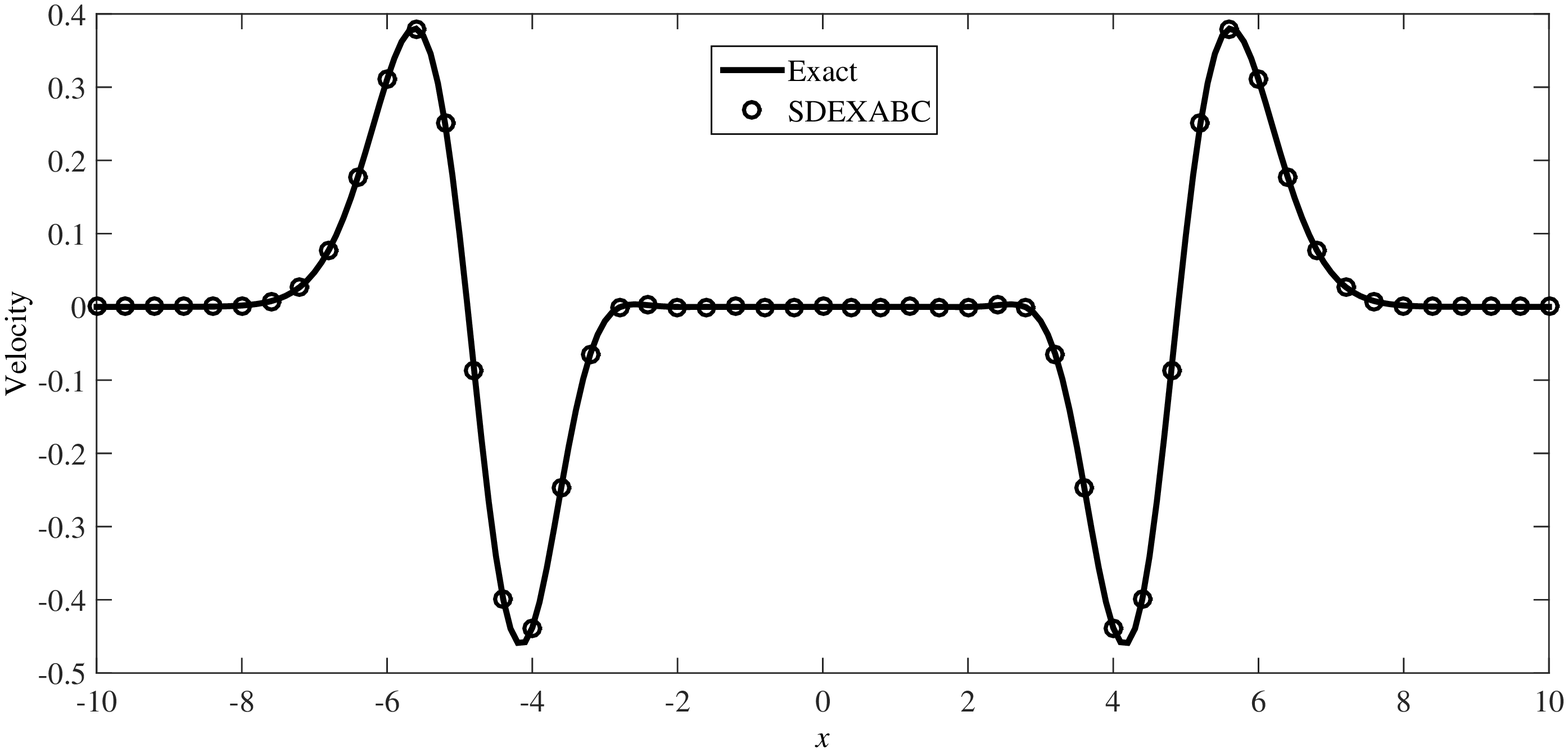}
\includegraphics[width=8cm]{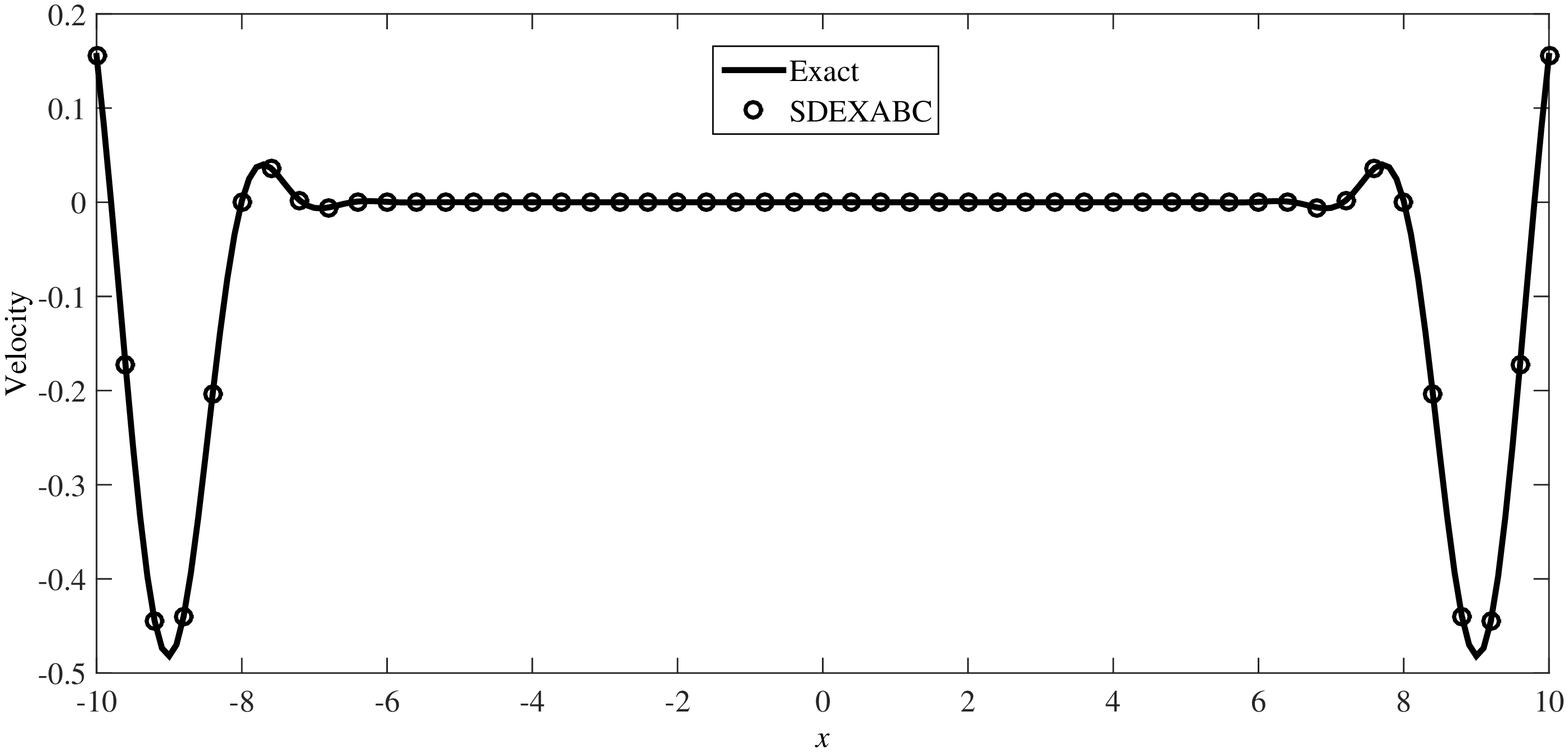}
\includegraphics[width=8cm]{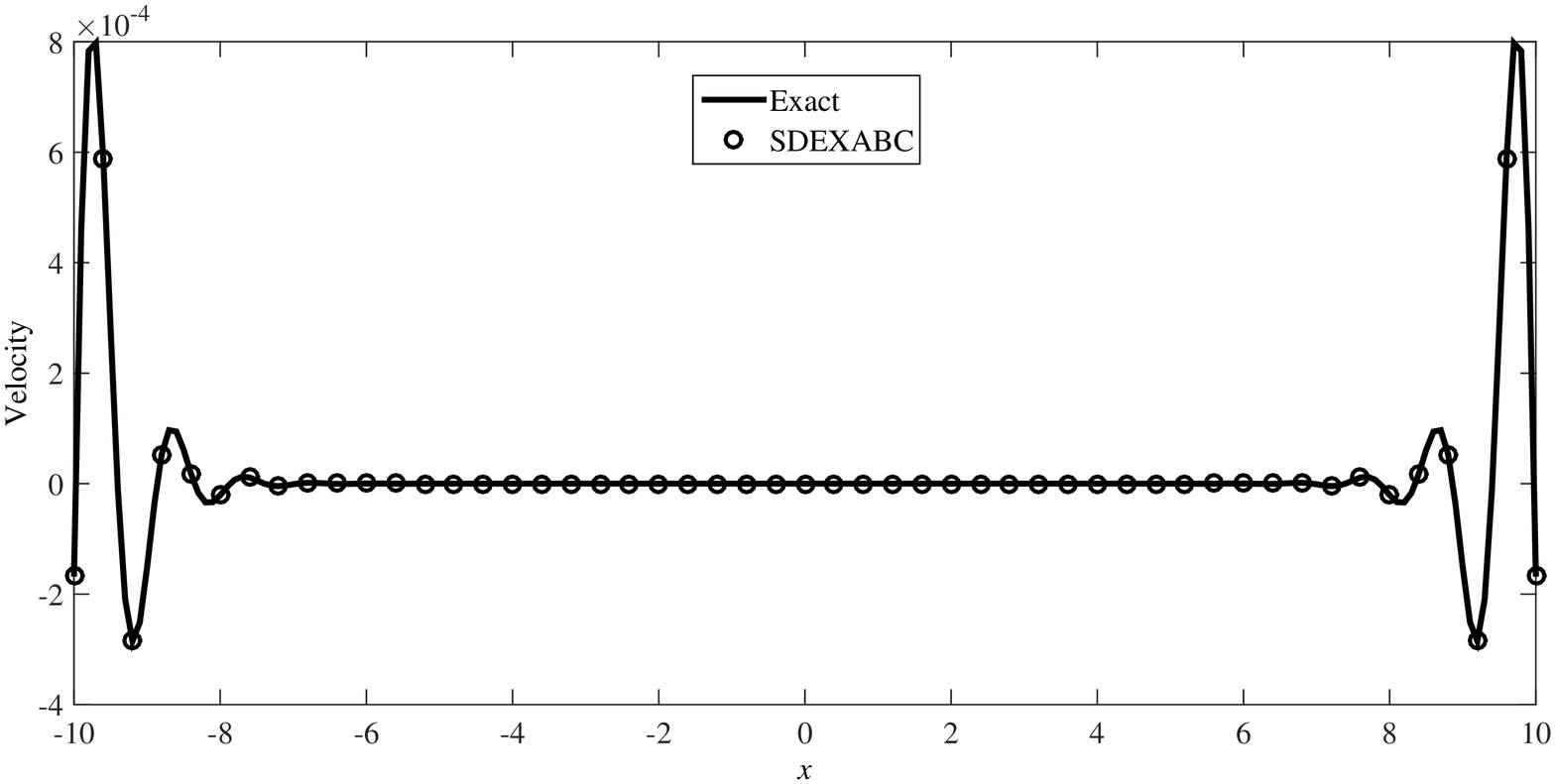}
\includegraphics[width=8cm]{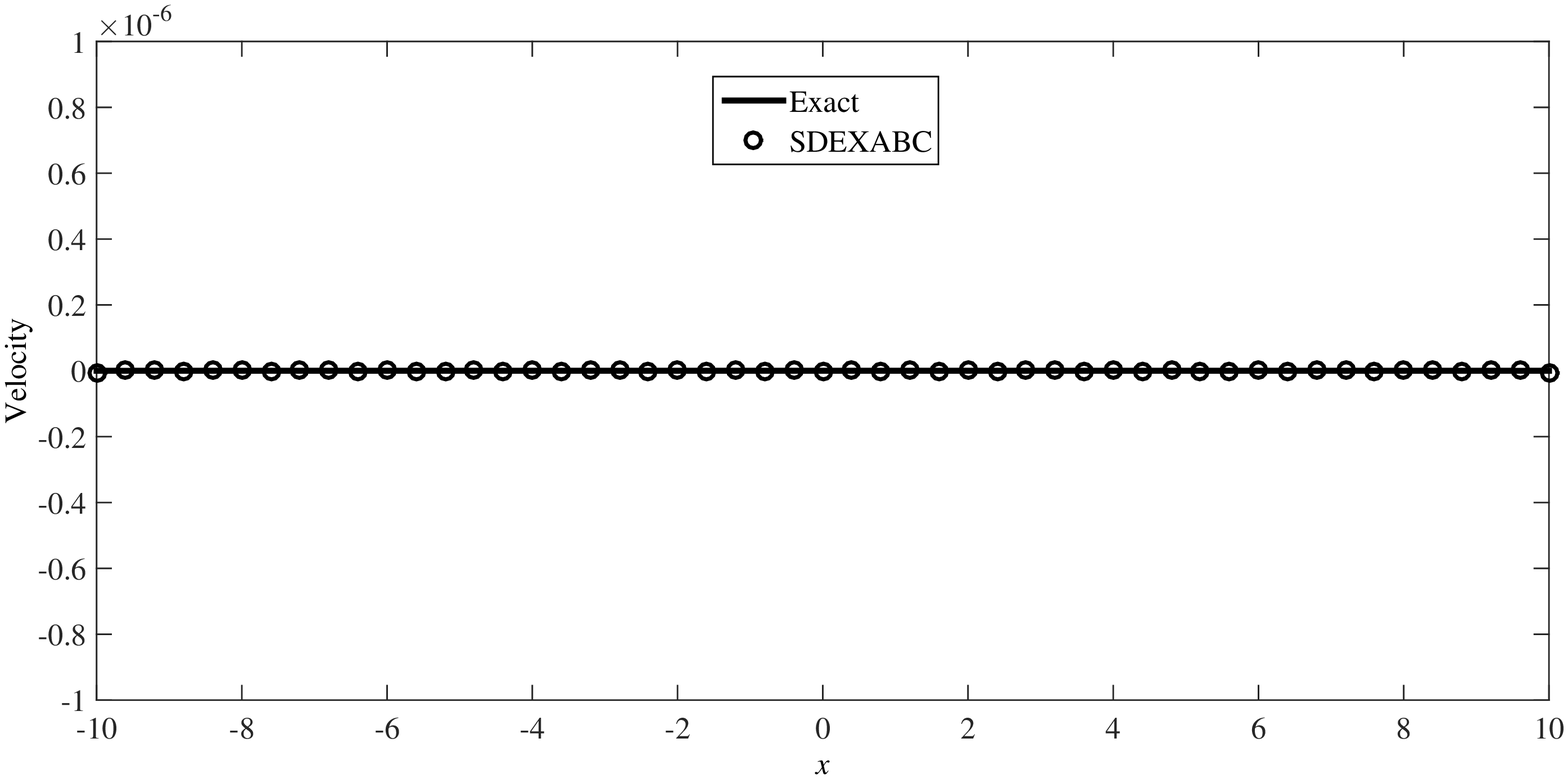}
\caption{Numerical solution computed: upper-left for $t=5$; upper-right for $t=10$; lower-left for $t=15$;
lower-right for $t=40$.}\label{vel}
\end{figure}

The numerical solutions of displacements with boundary condition (\ref{b1}) at different times in the truncated finite domain are plotted in Figure  \ref{bar}, along with the analytical solution(\ref{ana}). We can see the Gauss source splits into two humps and and then goes out of the computational domain in different directions. At $t=40$, the wave almost goes out of the computational domain. One can see a good agreement. The similar results of velocities and analytical solution of velocities at different times are displayed in Figure  \ref{vel}. In \cite{Linjuan} Wang did the computation until $T=15$, the numerical solution had deviated the analytical solution apparently at $t=15$, which illustrates the deficiency of non-exact boundary conditions.

\begin{figure}\centering
\includegraphics[width=8cm]{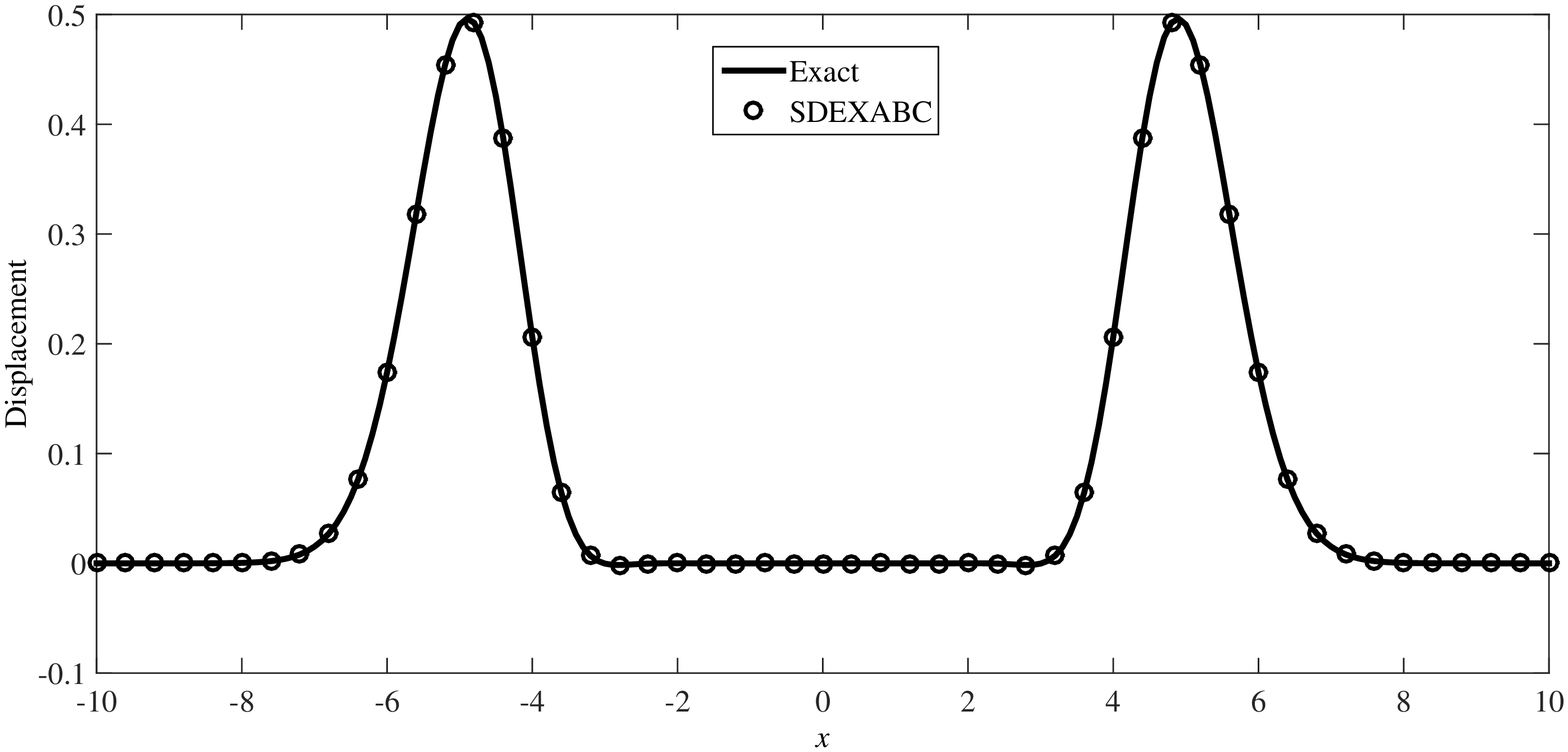}
\includegraphics[width=8cm]{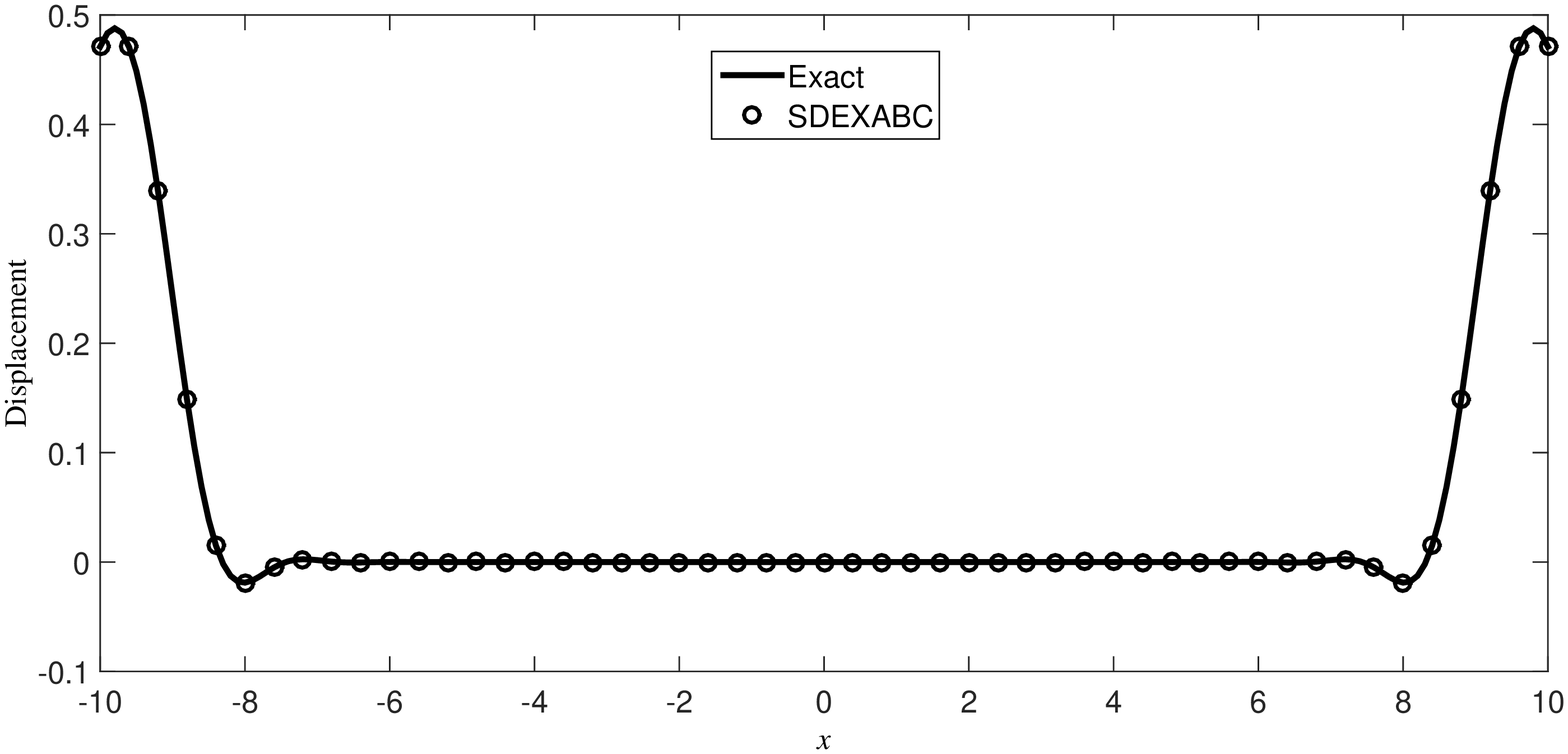}
\includegraphics[width=8cm]{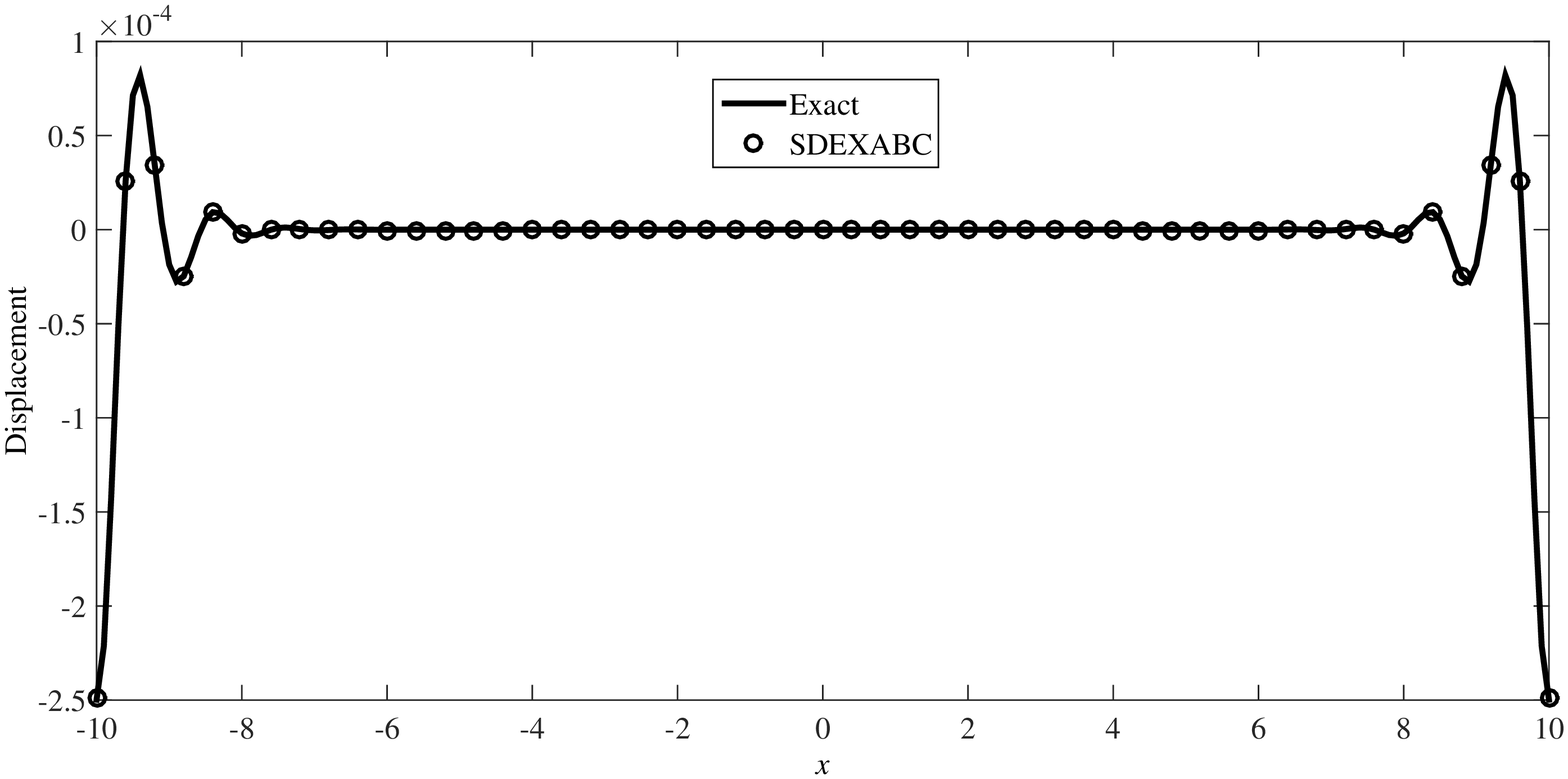}
\includegraphics[width=8cm]{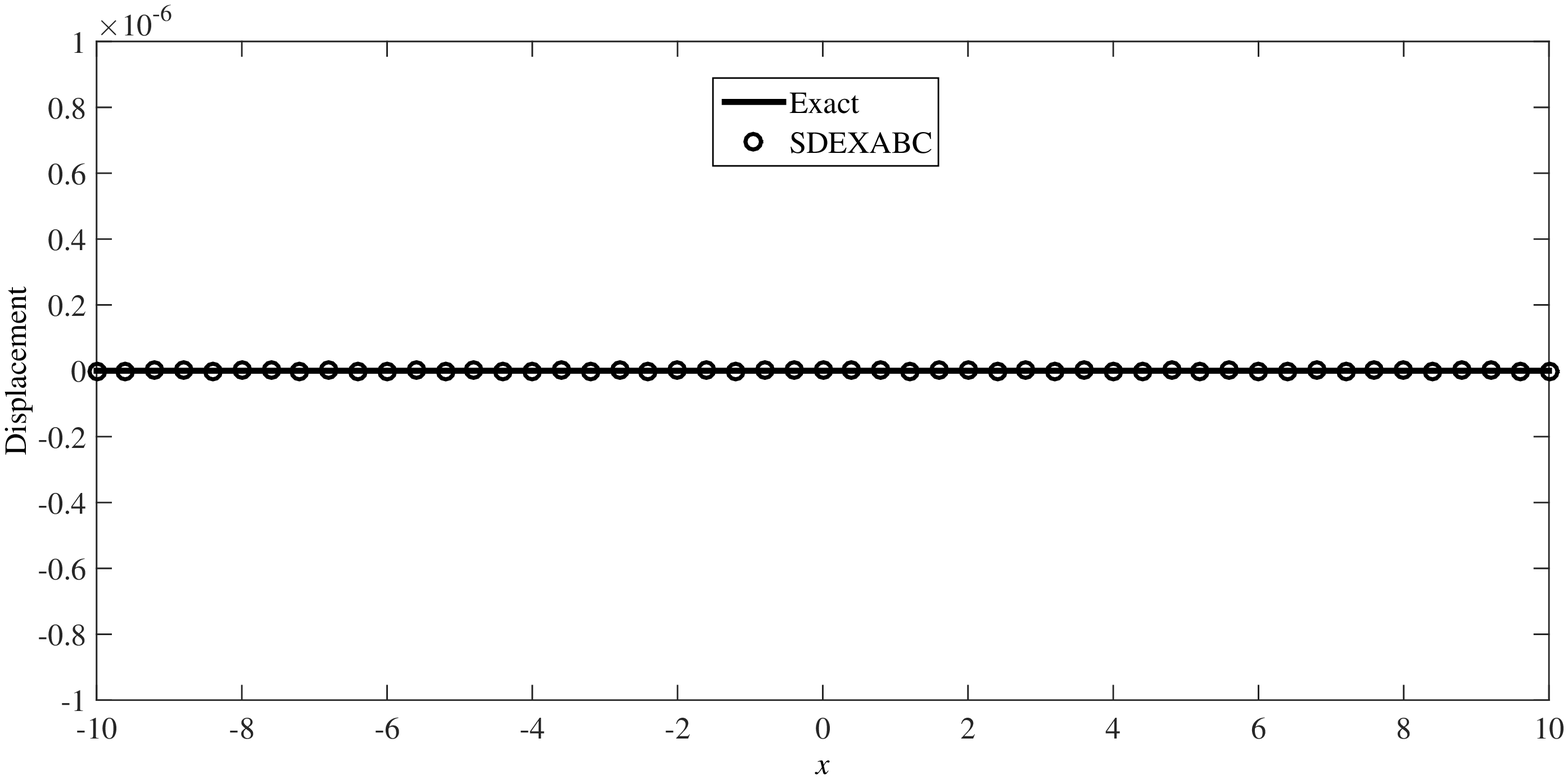}
\caption{Numerical solution computed: upper-left for $t=5$; upper-right for $t=10$; lower-left for $t=15$;
lower-right for $t=40$.}\label{bar}
\end{figure}

\begin{figure}\centering
\includegraphics[width=8cm]{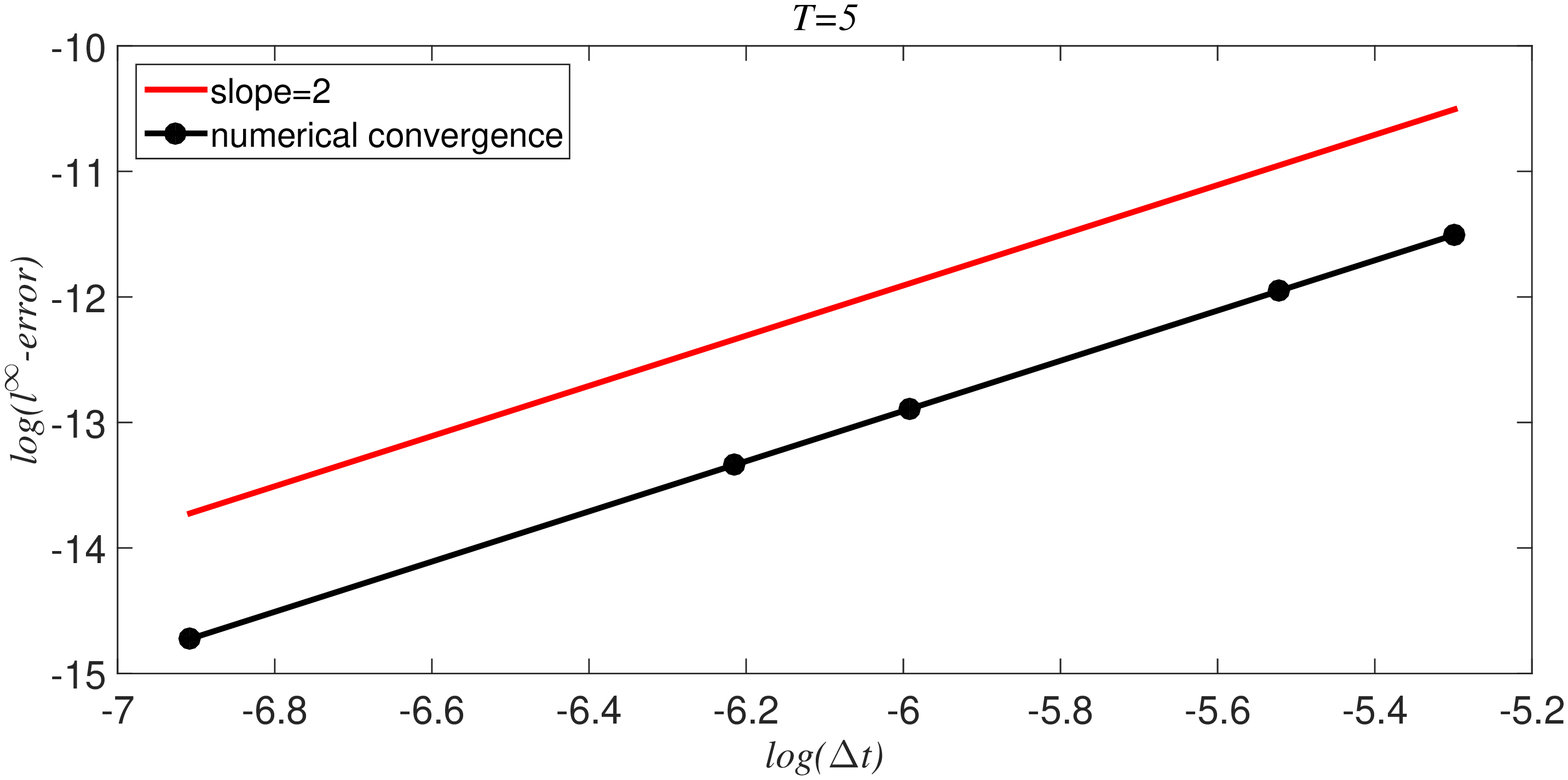}
\includegraphics[width=8cm]{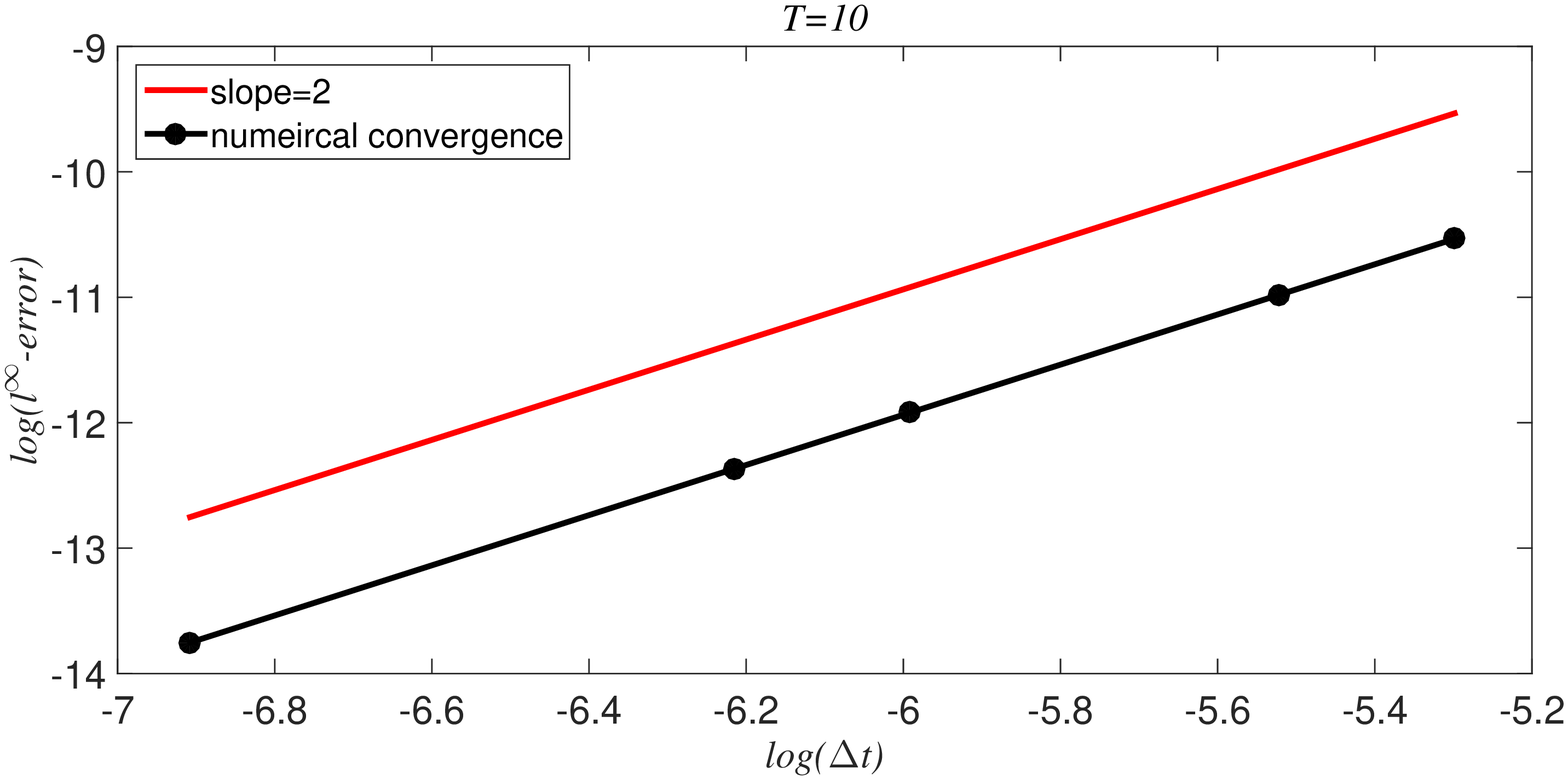}
\includegraphics[width=8cm]{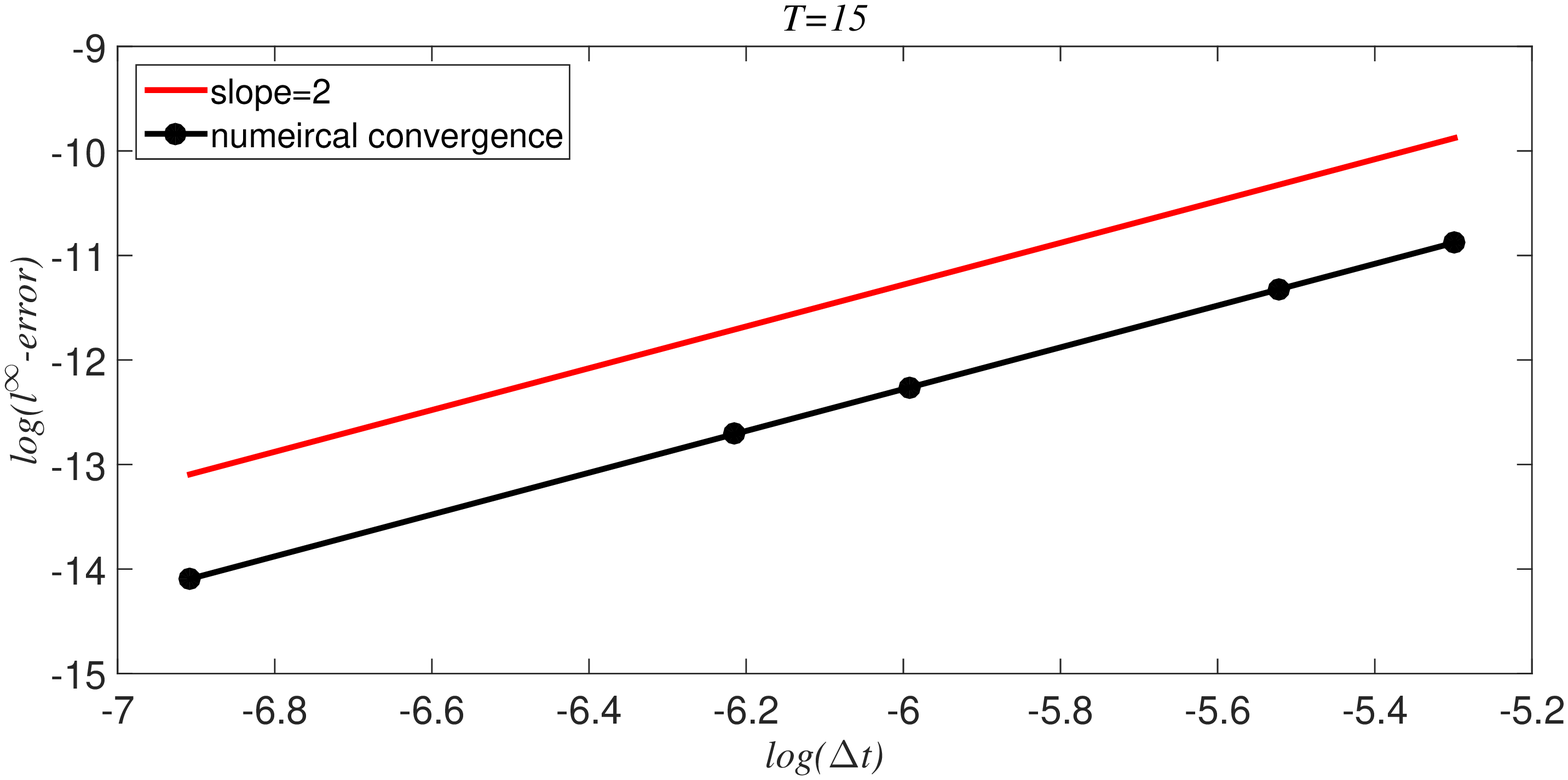}
\includegraphics[width=8cm]{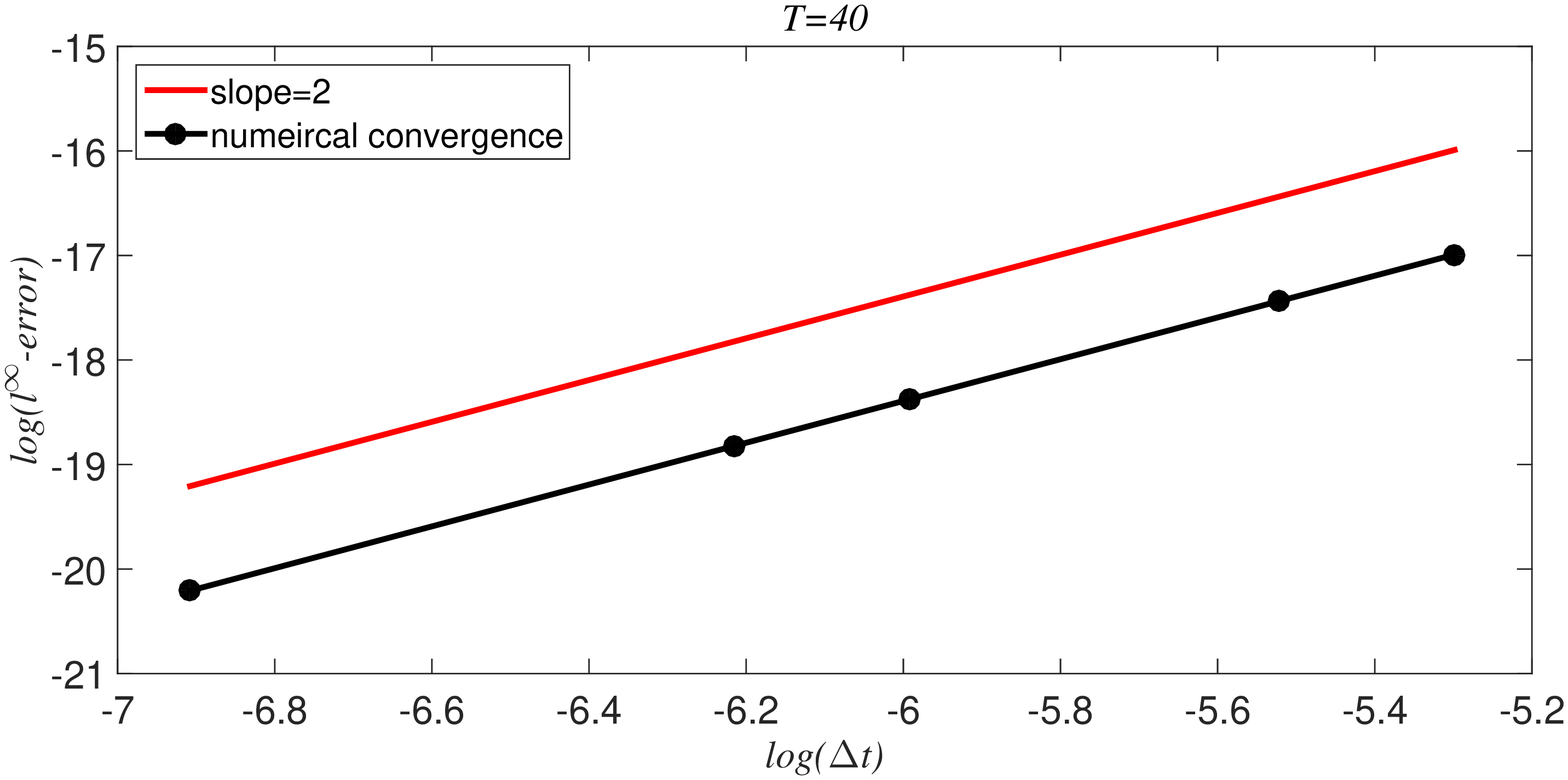}
\caption{Convergence rate of displacement at different time.}\label{uepic}
\end{figure}

The $l_{\infty}$-error of displacement \[\displaystyle\max_{1\leq n\leq 201}|u^{ref}_{n}-u_{n}|,\]
is used to measure the performance. The reference solution is obtained with a computational domain large enough such that these are no boundary reflections. The $l^{\infty}$ error is listed in Tables \ref{tabeuler1}. The convergence rates at different times are around second order. The convergence rates of displacement are depicted in Fig \ref{uepic}.
\begin{table}[h]
\begin{center}
\caption { $l_{\infty}$-error and convergence rate of displacement for different time.}\label{tabeuler1}
\begin{tabular}{|c|c|c|c|c|c|c}\hline
$l_{\infty}$-error &$\Delta t=0.001$&$\Delta t=0.002$&$\Delta t=0.0025$&$\Delta t=0.004$& $\Delta t=0.005$&rate \\\hline

$t=5$    & 4.03e-07 &  1.61e-06 & 2.51e-06  & 2.51e-06 & 1.01e-05 &2.00\\\hline
$t=10$    & 1.06e-06 &  4.26e-06 & 6.65e-06  & 1.70e-05& 2.66e-05&2.00\\\hline
$t=15$     & 7.57e-07 &  3.02e-06 & 4.72e-06  & 1.21e-05 & 1.88e-05 &1.99 \\\hline
$t=40$    & 1.67e-09 &  6.69e-09 & 1.04e-08  & 2.67e-08 & 4.17e-08 &2.00\\\hline
\end{tabular}
\end{center}
\end{table}

\begin{figure}\centering
\includegraphics[width=8cm]{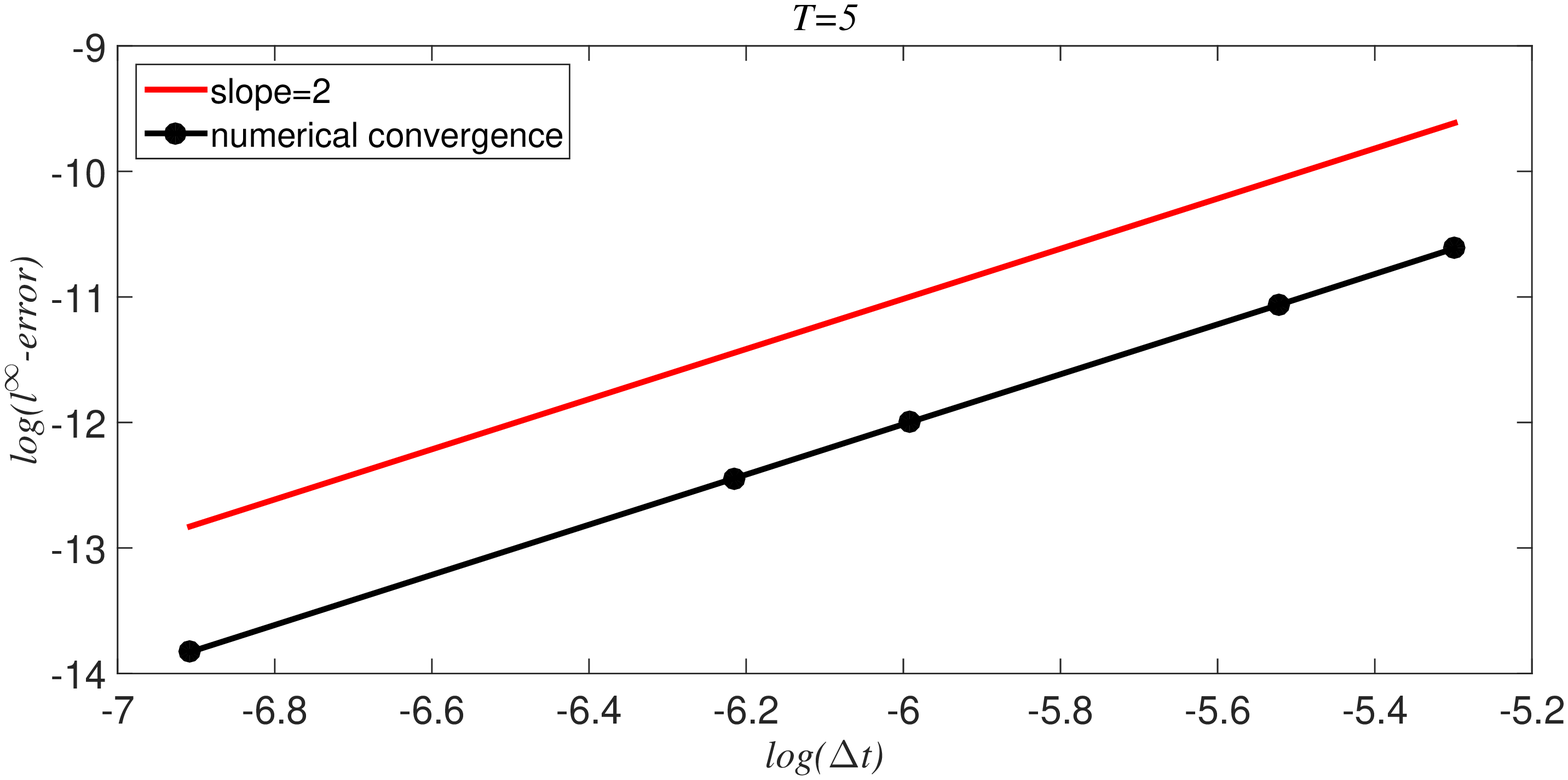}
\includegraphics[width=8cm]{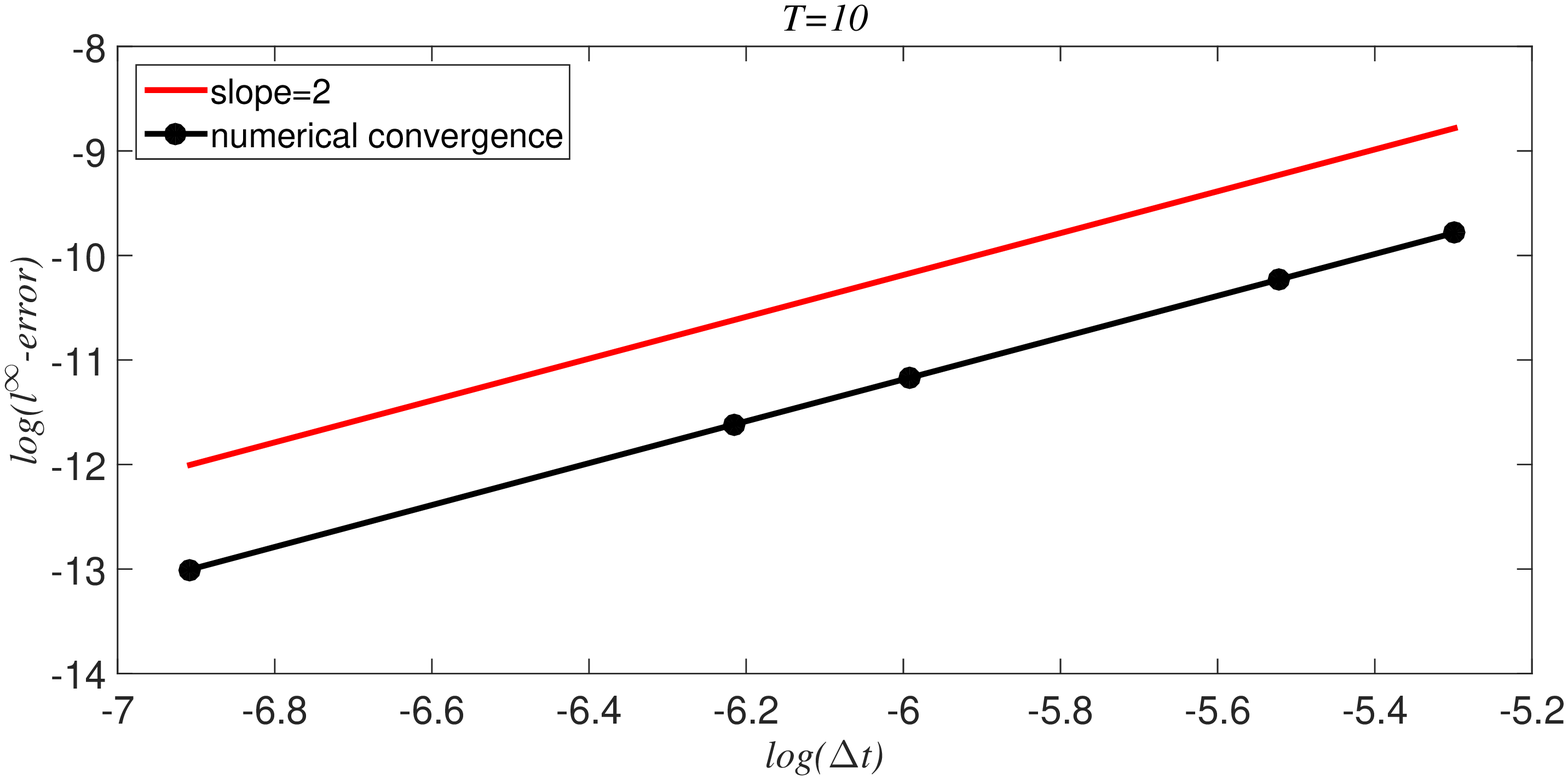}
\includegraphics[width=8cm]{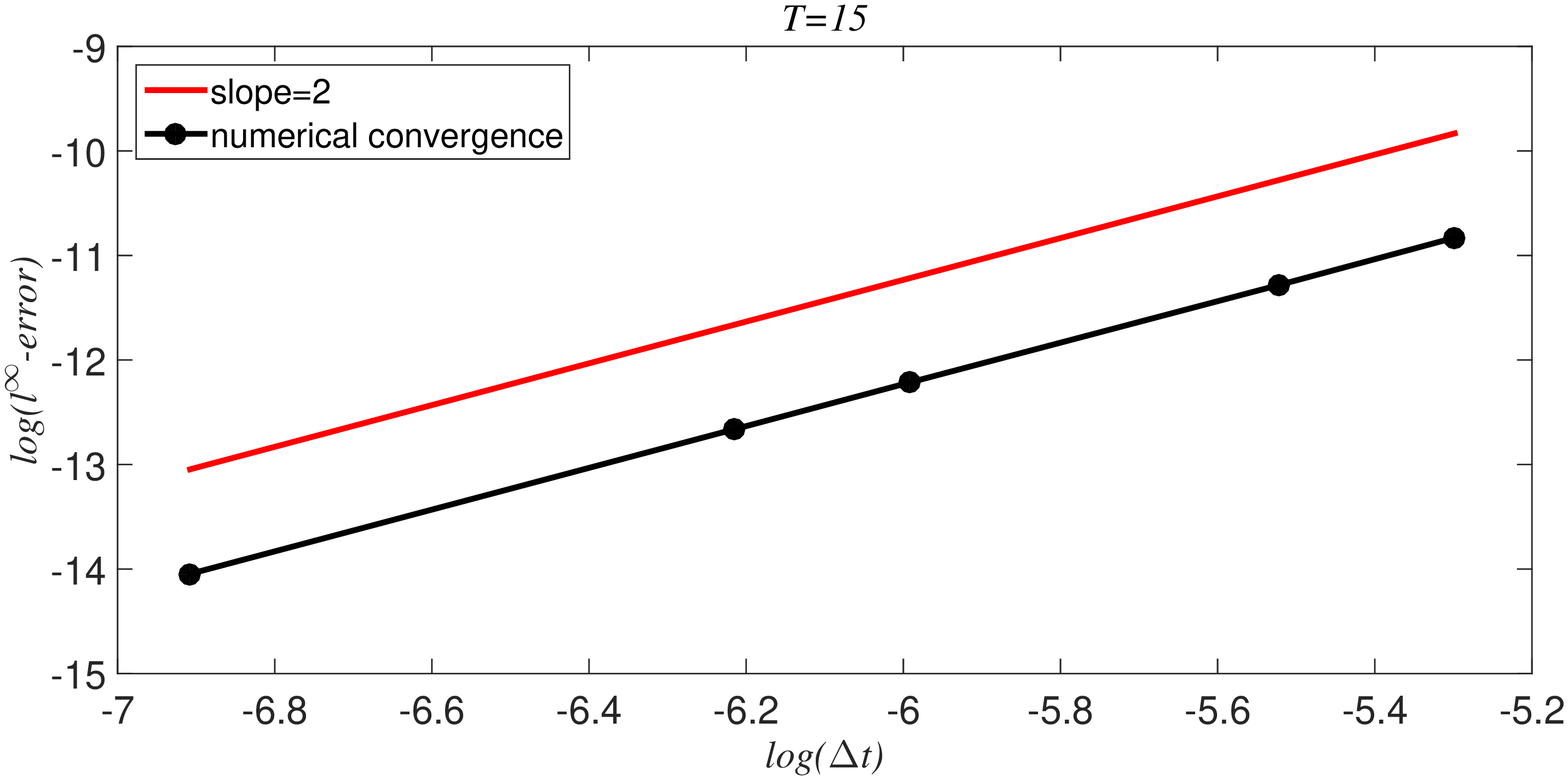}
\includegraphics[width=8cm]{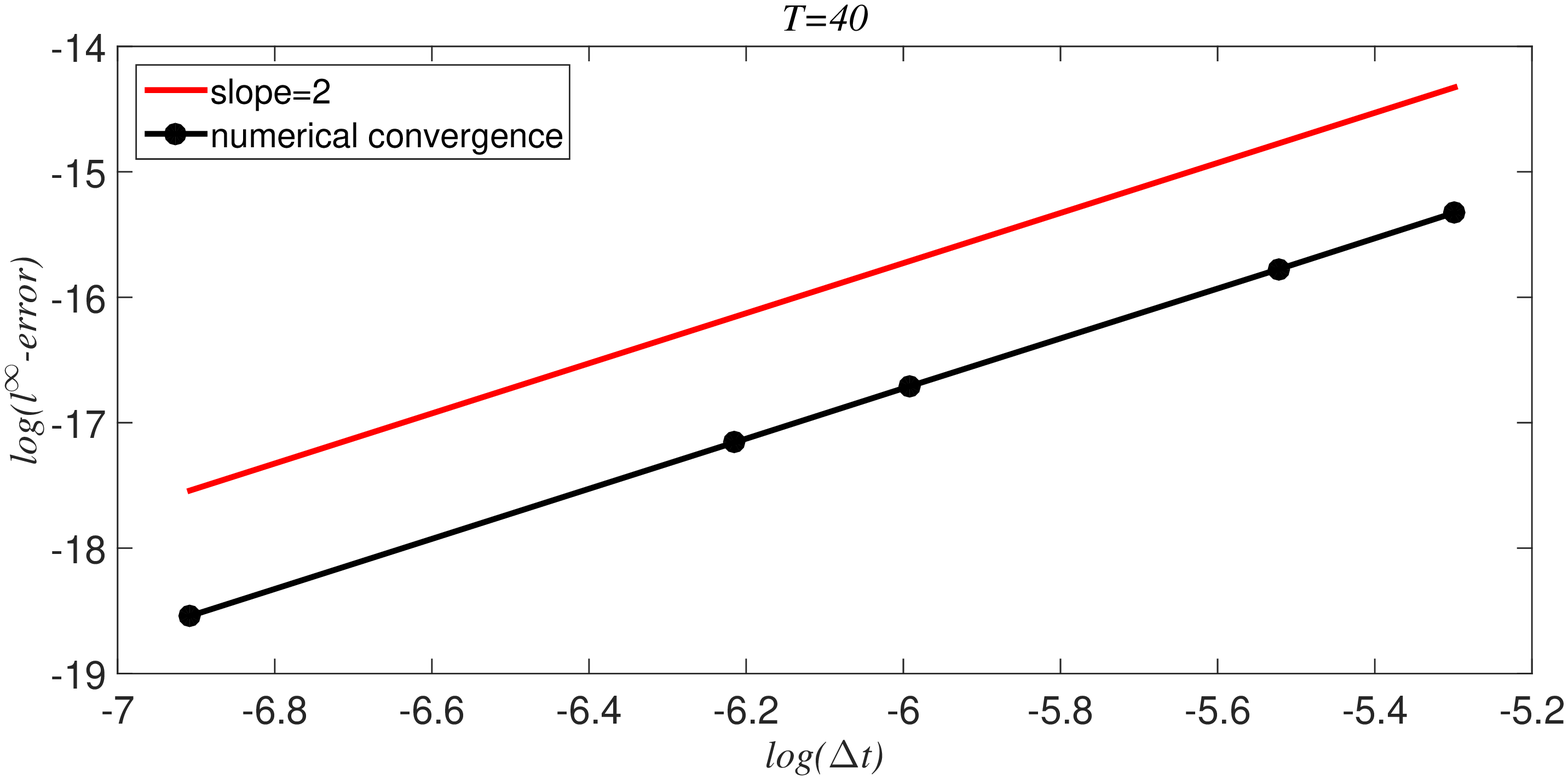}
\caption{Convergence rate of velocity at different time.}\label{vepic}
\end{figure}

In the same way, the $l^{\infty}$ errors of velocity are listed in Tables \ref{tabeuler2}. The convergence rates at different times are
again of second order. The convergence rates of velocity are depicted in Fig \ref{vepic}.
\begin{table}[h]
\begin{center}
\caption { $l_{\infty}$-error and convergence rate of velocity for different time.}\label{tabeuler2}
\begin{tabular}{|c|c|c|c|c|c|c}\hline
$l_{\infty}$-error &$\Delta t=0.001$&$\Delta t=0.002$&$\Delta t=0.0025$&$\Delta t=0.004$& $\Delta t=0.005$&rate \\\hline

$t=5$    & 9.87e-07 &  3.94e-06 & 6.16e-06  & 1.57e-05 & 2.46e-05 &2.00\\\hline
$t=10$    & 2.25e-06 &  9.03e-06 & 1.41e-05  & 3.61e-05& 5.64e-05&2.00\\\hline
$t=15$     & 7.94e-07 &  3.17e-06 & 4.96e-06  & 1.26e-05 & 1.98e-05 &2.00 \\\hline
$t=40$    & 8.87e-09 &  3.54e-08 & 5.54e-08  & 1.41e-07 & 2.21e-07 &2.00\\\hline
\end{tabular}
\end{center}
\end{table}

\subsection{Interface}\label{Inter}
\begin{figure}\centering
\includegraphics[width=8cm]{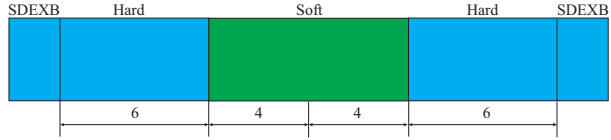}
\caption{A segment of an infinite composite bar with interfaces.}\label{hard}
\end{figure}

We next study propagation of a Gauss source in an infinite bar which is composed of different materials,
the truncated computational domain can be chosen as $[-10,10]$ shown in Figure \ref{hard} just as \cite{Linjuan}. In the central green domain of a length 8 the Young's moduli is $E'$ and on the blue domain the Young's moduli is $E$ where the Young's moduli satisfies $E'=\beta E$. The $\beta <1$ means the Young's moduli of material in the green domain is smaller than the that of the material on the blue domain. Therefore the green material is softer, we denote the green domain as $D_{soft}$. In the same way we denote the materials on the two sides as $D_{hard}$. The two materials have the same period horizon $\delta$, and the micromodulus function composite bar can be taken as

\begin{eqnarray}C_{bar}=
\begin{cases}
C(x'-x), &x'\in D_{hard}, x\in D_{hard},  \cr \beta C(x'-x), & \text{otherwise,}
\end{cases}
\end{eqnarray}
where $C$ can be taken as (\ref{modu}). $\delta$ have the same setting as section \ref{bar}.

\begin{figure}\centering
\includegraphics[width=10cm]{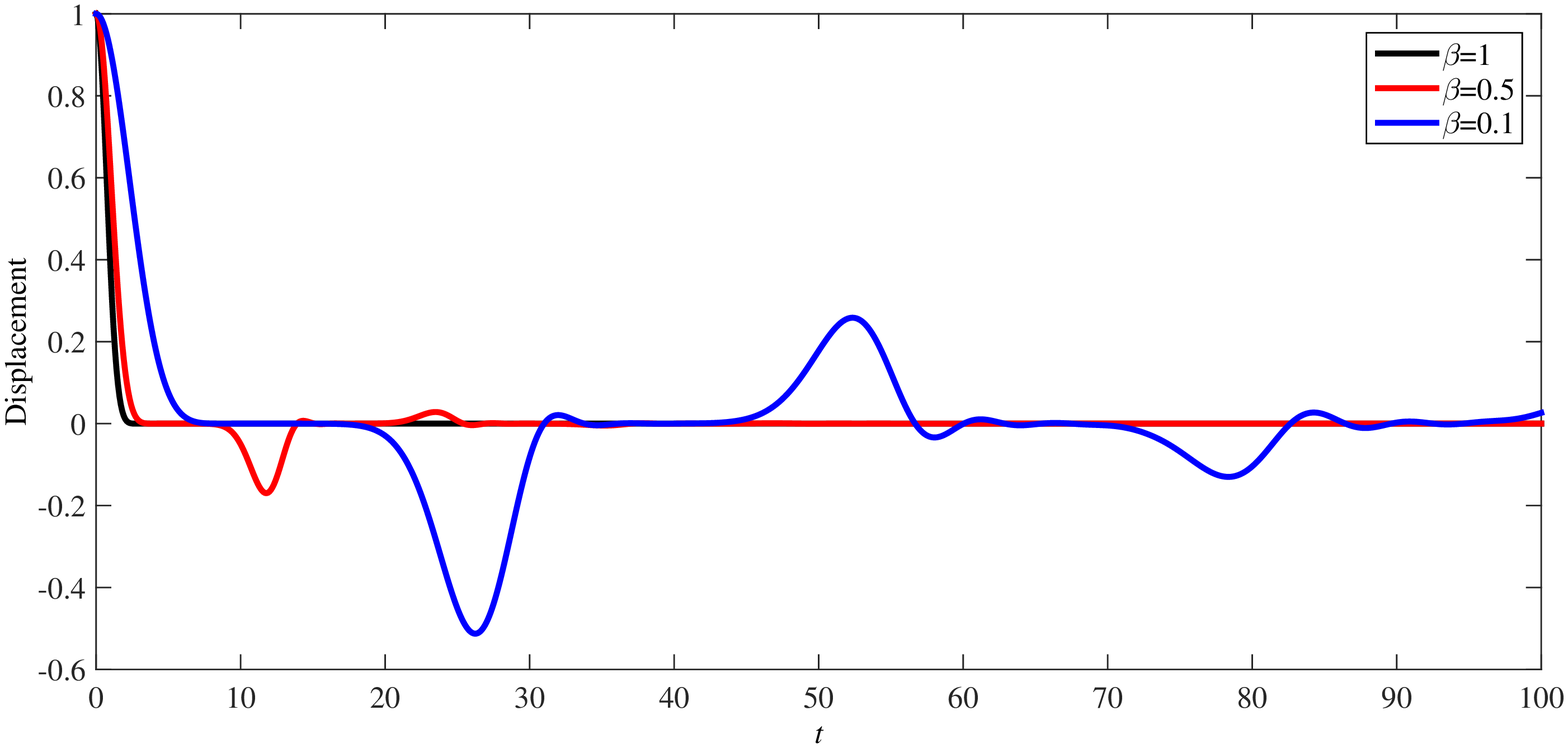}
\includegraphics[width=10cm]{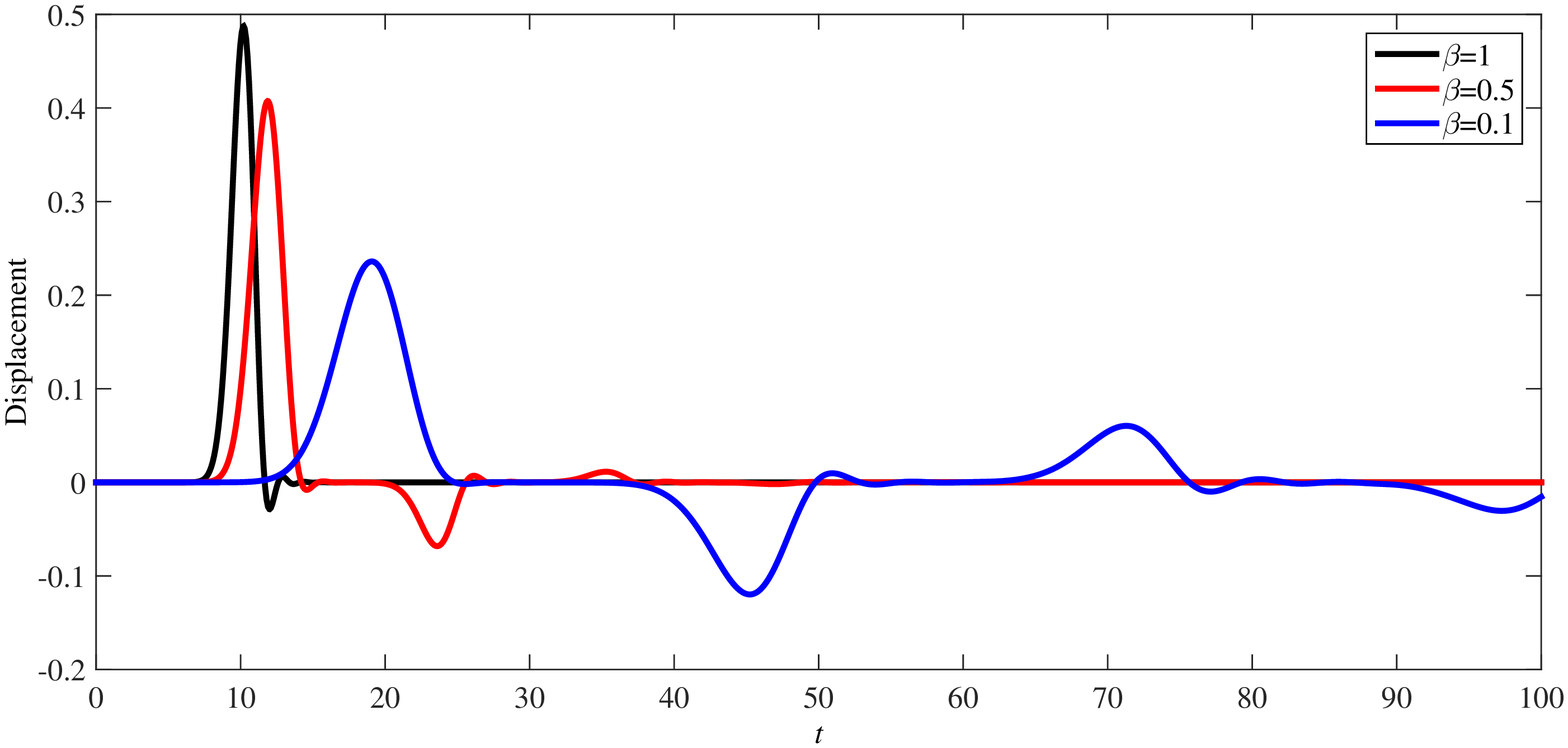}
\caption{Displacement for evolution with respect to $t$ (a) the origin where all peaks are due to reflection waves by the
interface (except that at t D 0); (b) the right end where all peaks are due to transmission waves.}\label{fdmr}
\end{figure}

Under initial conditions (\ref{b1}), the displacements for different values of locations are plotted in Figure \ref{fdmr}. Because of symmetry we only plot the displacements of the origin in Figure \ref{fdmr} (a) and the right end in Figure \ref{fdmr} (b), respectively. From the Figure \ref{fdmr} it is clear that different $\beta$s bring different velocities and different size of the humps and hollows. When $\beta=1$, this example is just same as the test in Section \ref{bar}. We see no humps or hollows as the origin evolves.  If $\beta<1$, the splitting humps of the origin first reach the interfaces of the two materials and then generate reflection
and transmission at the interfaces. As $\beta$ decreases, the difference of the two material Young's modulis grows bigger, the amplitudes of the reflection wave increase and those of the transmission
wave decrease. On the other hand the reflection waves have the opposite vibration direction with the transmission wave. In Figure \ref{fdmr} (b) the first hump of blue line and that of red line represents transmission wave at the interface, therefor we see in Figure \ref{fdmr} (b) the first hump of blue line is smaller than that of red line. In the same way we see in Figure \ref{fdmr} (a) the first hollow of blue line is bigger than that of red line because that the reflection
wave increases as the $\beta$ decreases when the splitting humps of the origin first reach interface. Then
the reflection waves pass through each other and reach the interface, the reflection wave will generate new reflection wave at the interface, we can see the reflection wave in the second hump of blue line in Figure \ref{fdmr} (a).  The process goes on as before.

\begin{figure}\centering
\includegraphics[width=10cm]{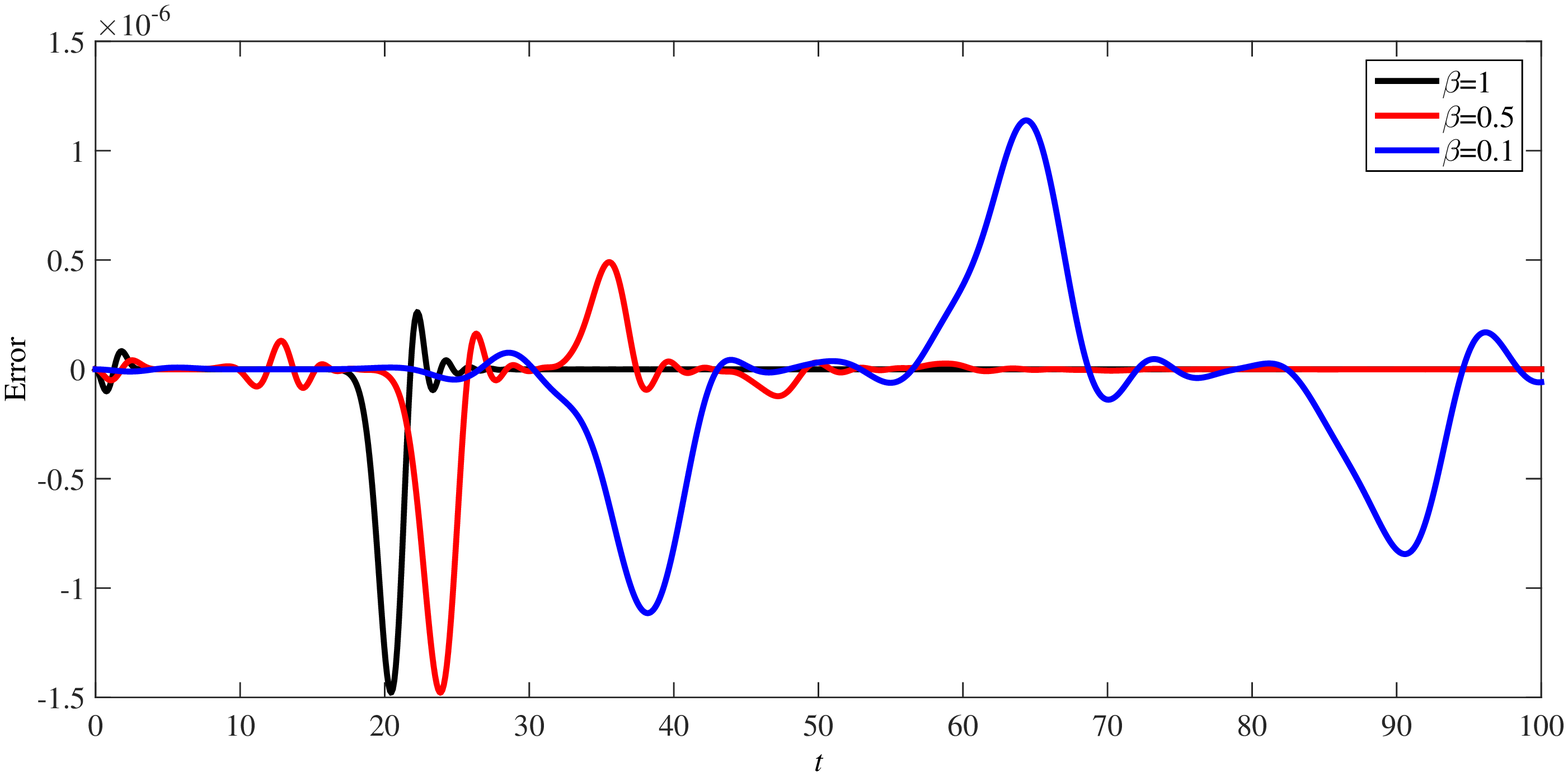}
\includegraphics[width=10cm]{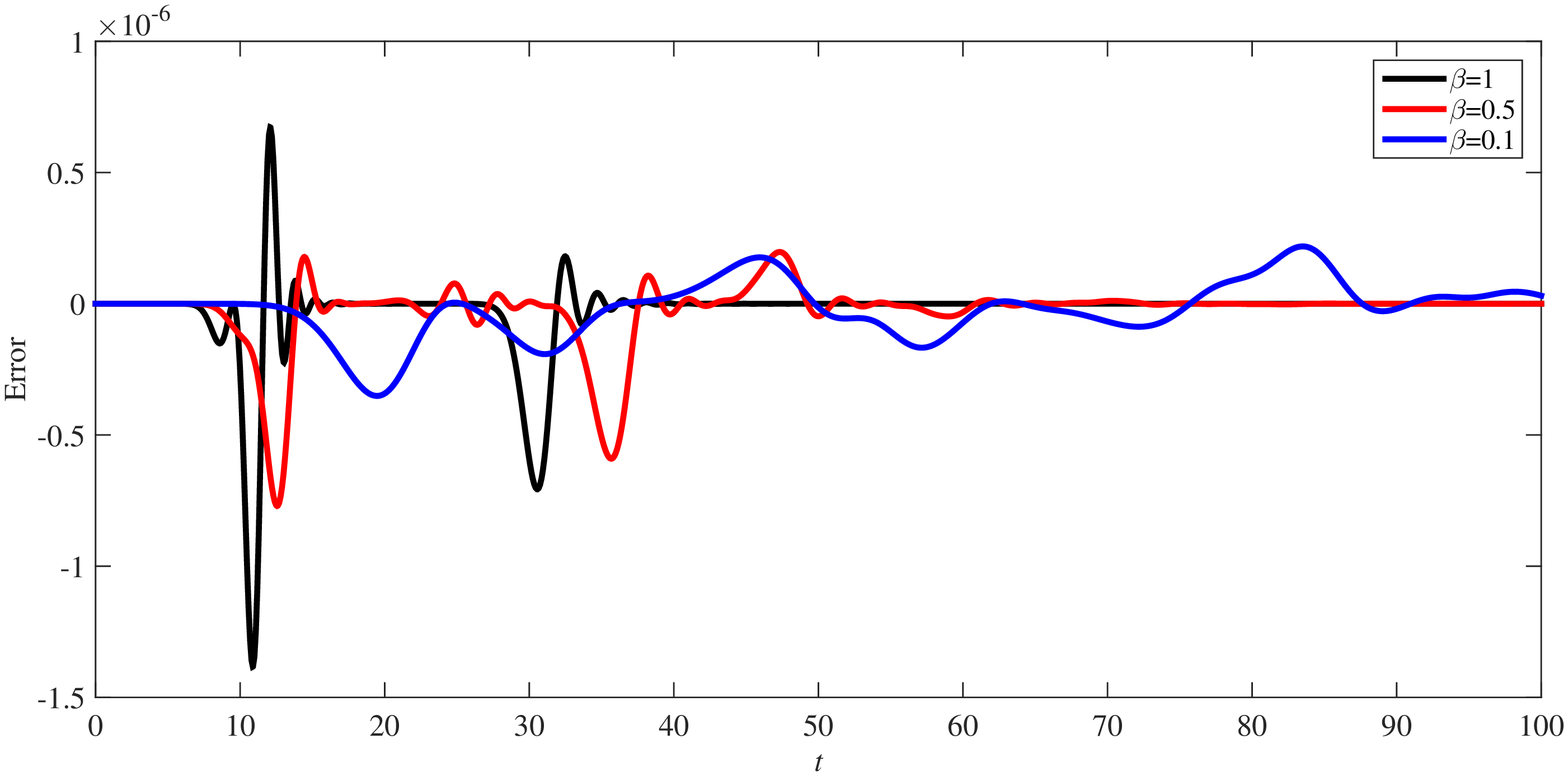}
\caption{Error for evolution with respect to $t$: (a) evolution of error for origin; (b) evolution of error for right end.}\label{error}
\end{figure}

The errors of displacements between reference solution and numerical solution are plotted in Figure \ref{error}. One can see a good agreement.

\begin{figure}\centering
\includegraphics[width=8cm]{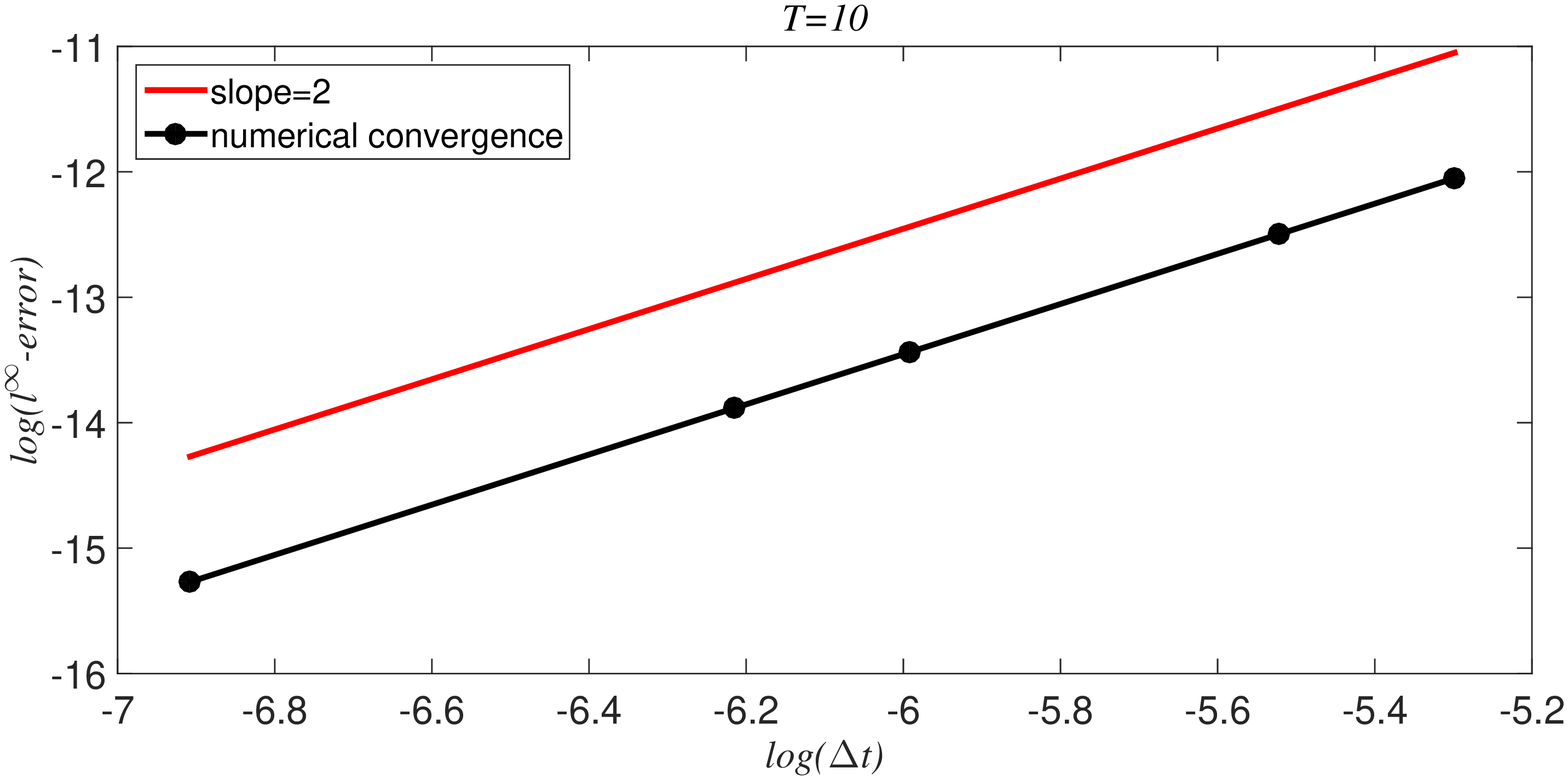}
\includegraphics[width=8cm]{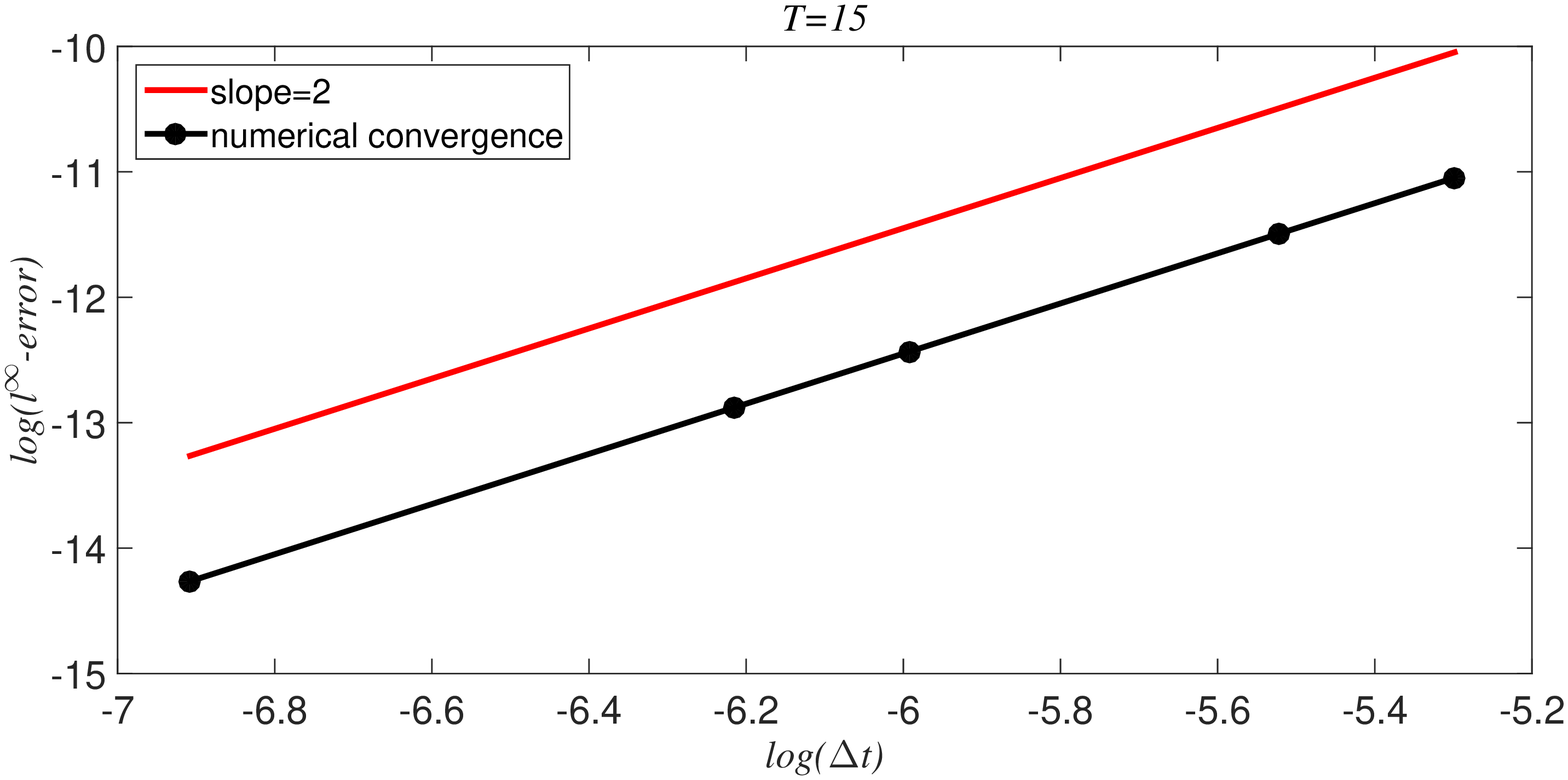}
\includegraphics[width=8cm]{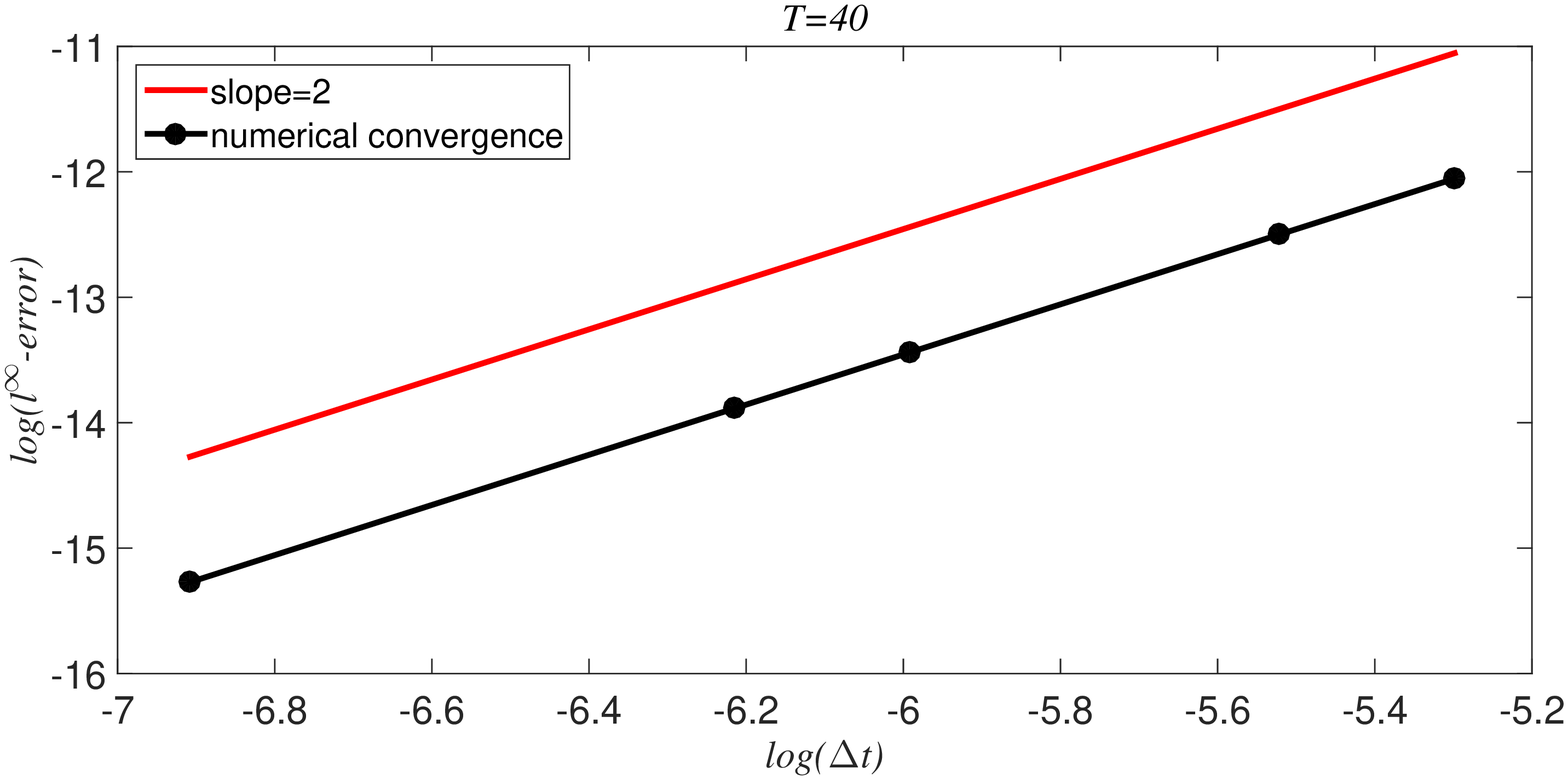}
\includegraphics[width=8cm]{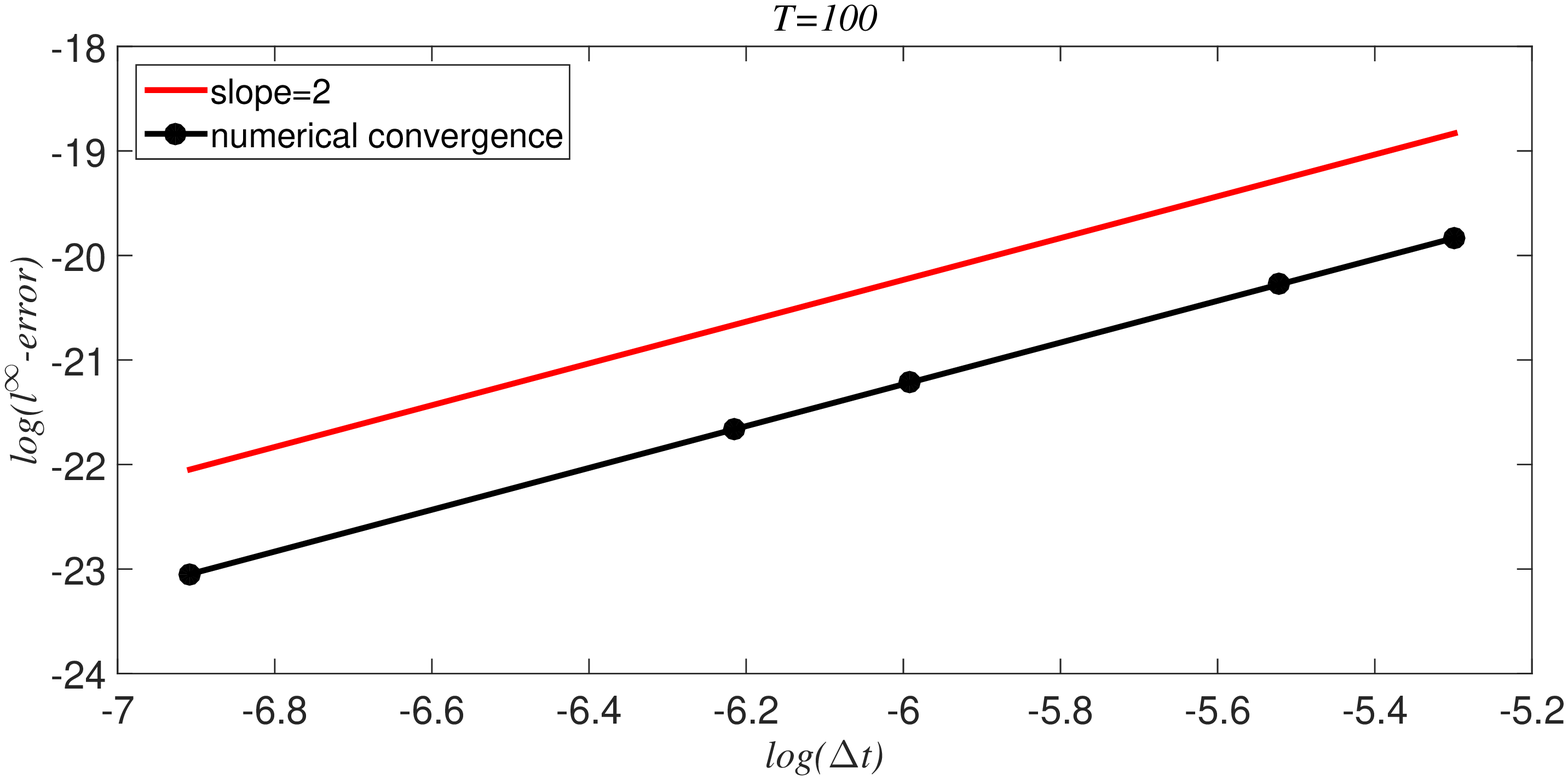}
\caption{Convergence rate of displacement at different time with $\beta=0.5$.}\label{ue0.5prob2}
\end{figure}

\begin{figure}\centering
\includegraphics[width=8cm]{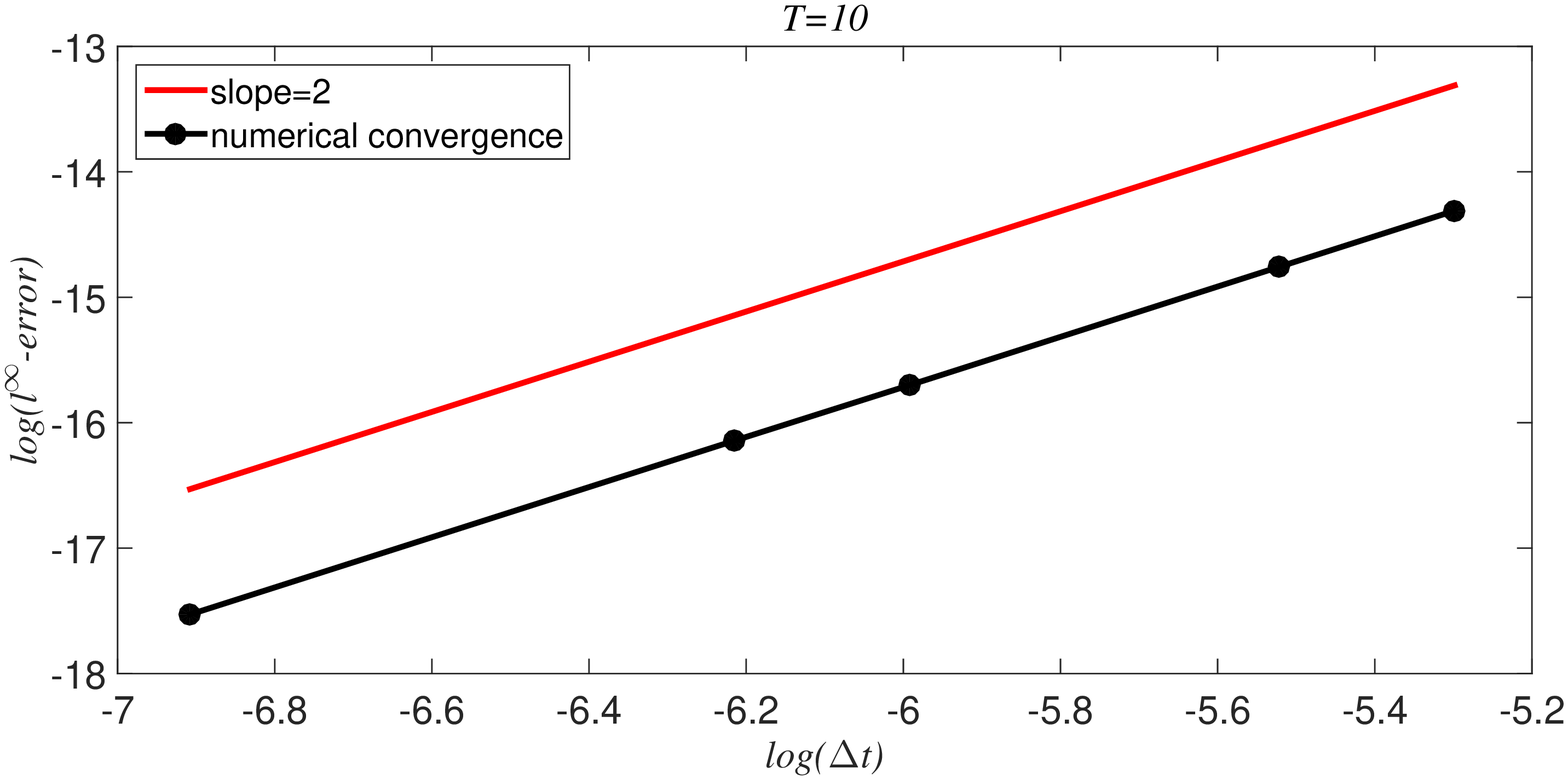}
\includegraphics[width=8cm]{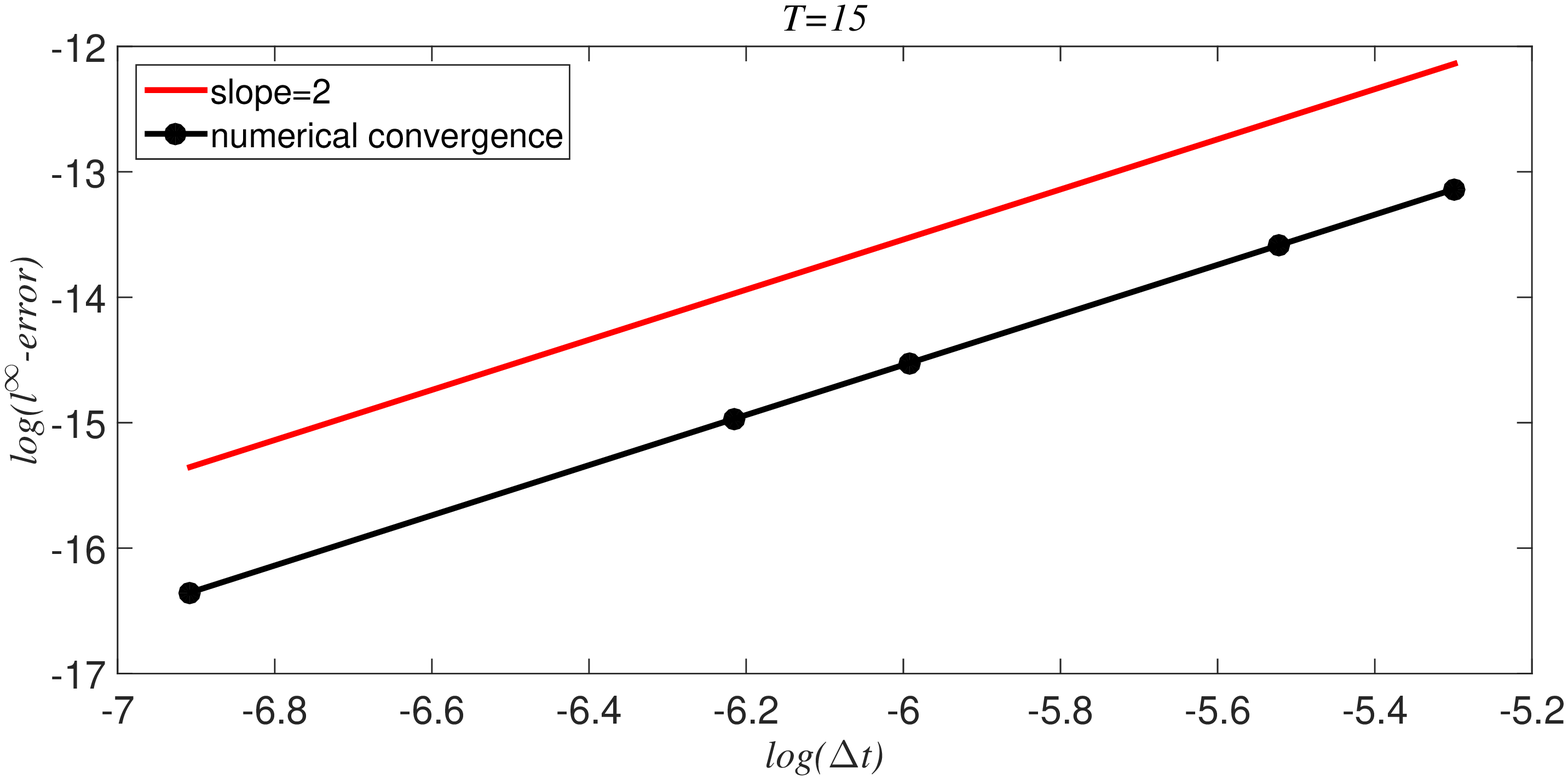}
\includegraphics[width=8cm]{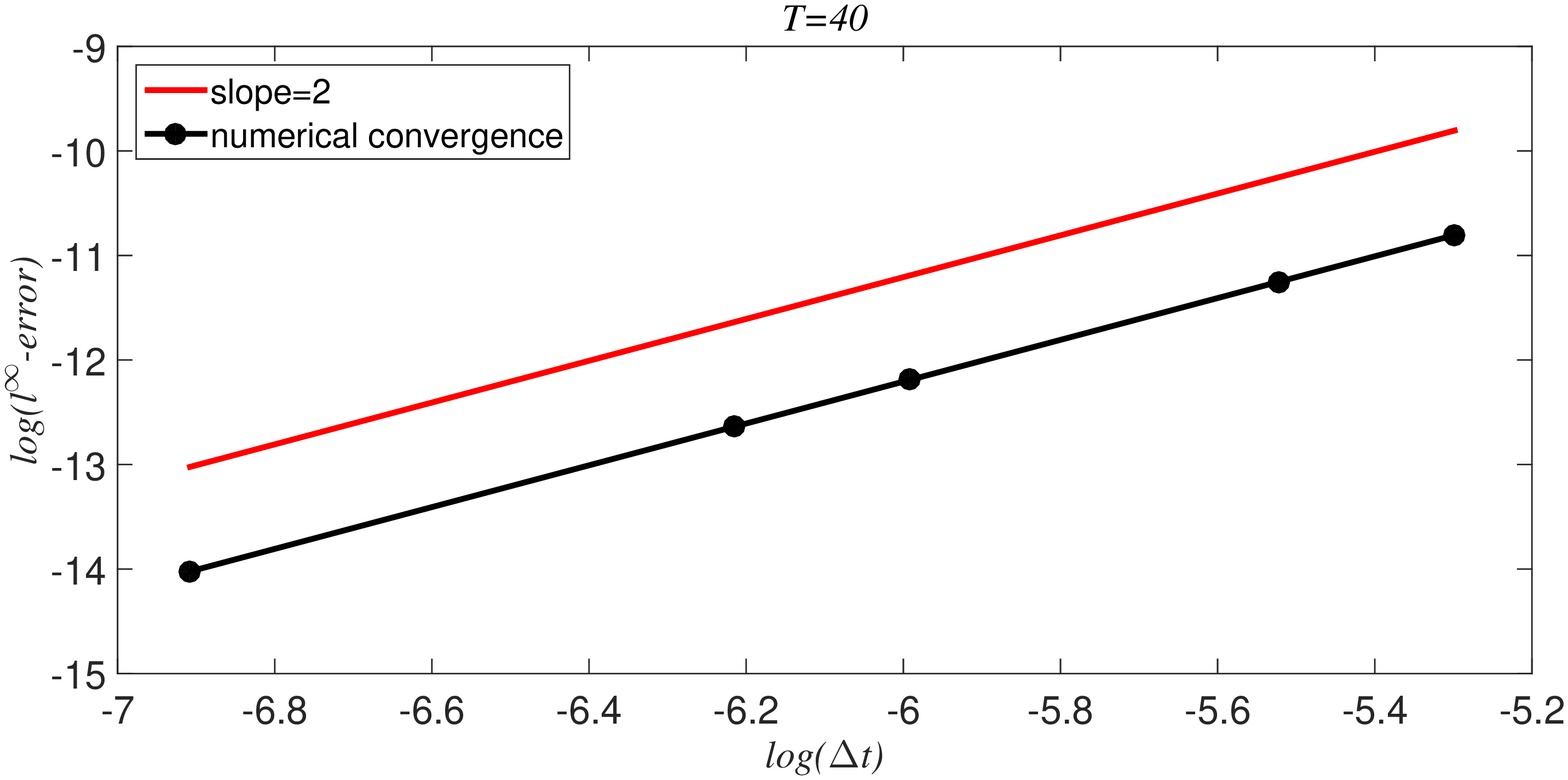}
\includegraphics[width=8cm]{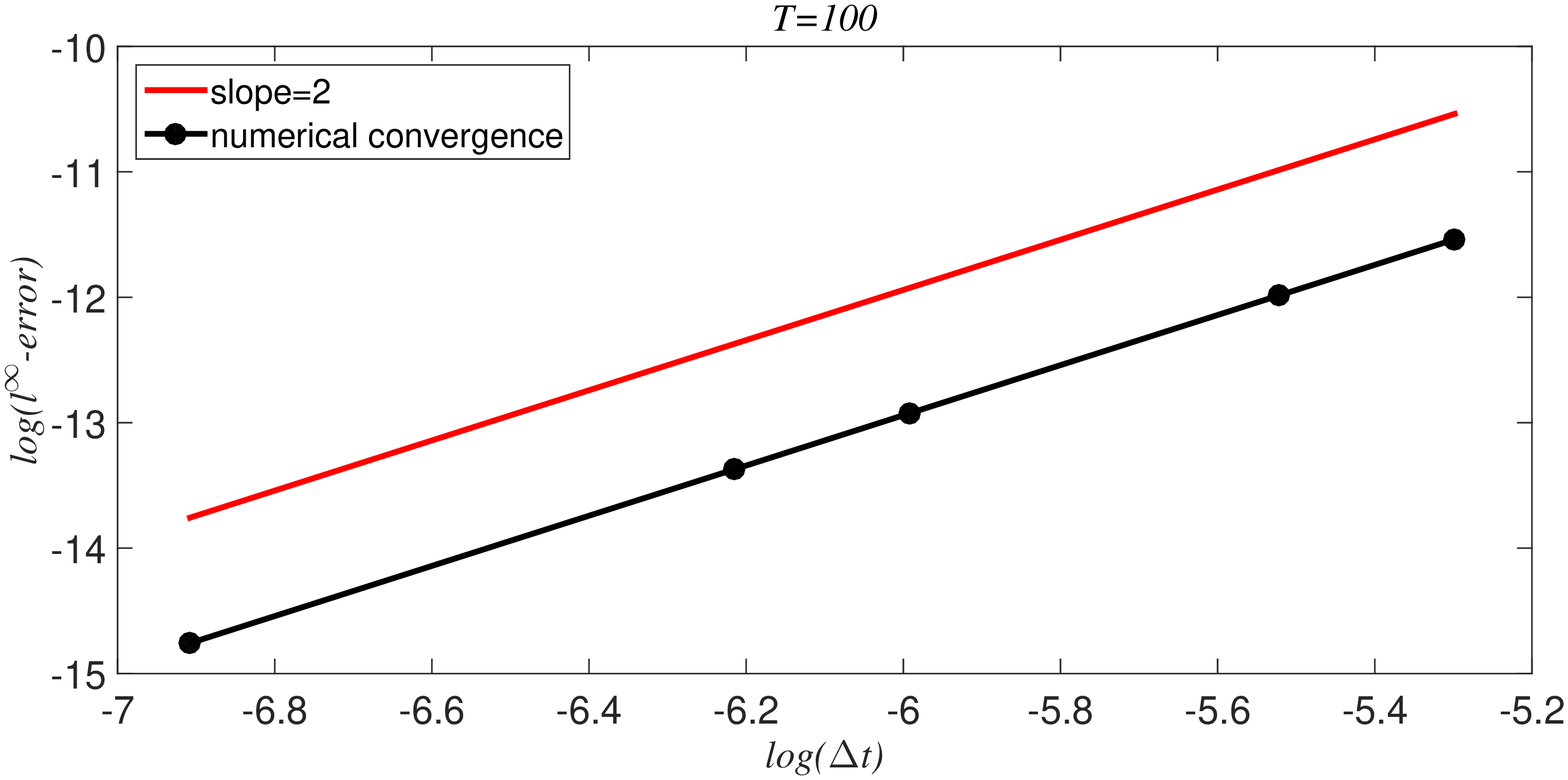}
\caption{Convergence rate of displacement at different time with $\beta=0.1$.}\label{ueprob2}
\end{figure}

The $l^{\infty}$-errors with $\beta=0.5$ and $\beta=0.1$ are listed in Tables \ref{fenduan0.5} and  \ref{fenduan}, respectively. The convergence rates at different times are again of second order. The convergence rates of displacement with $\beta =0.5$ and $\beta =0.1$ are depicted in Figures \ref{ue0.5prob2} and \ref{ueprob2}, respectively.

\begin{table}[h]
\begin{center}
\caption { $l_{\infty}$-error and convergence rate of displacement with $\beta=0.5$ for different time.}\label{fenduan0.5}
\begin{tabular}{|c|c|c|c|c|c|c}\hline
$l_{\infty}$-error &$\Delta t=0.001$&$\Delta t=0.002$&$\Delta t=0.0025$&$\Delta t=0.004$& $\Delta t=0.005$&rate \\\hline

$t=10$    & 2.33e-07 &  9.34e-07 & 1.46e-06  & 3.73e-06 & 5.84e-06 &2.00\\\hline
$t=15$    & 6.39e-07 &  2.55e-06 & 3.99e-06  & 1.02e-05& 1.59e-05&2.00\\\hline
$t=40$     & 2.33e-07 &  9.33e-06 & 1.45e-06  & 3.72e-06 & 5.82e-06 &2.00 \\\hline
$t=100$    & 9.78e-11 &  3.9e-10 & 6.11e-10  & 1.56e-09 & 2.43e-09 &2.00\\\hline
\end{tabular}
\end{center}
\end{table}

\begin{table}[h]
\begin{center}
\caption { $l_{\infty}$-error and convergence rate of displacement with $\beta=0.1$ for different time.}\label{fenduan}
\begin{tabular}{|c|c|c|c|c|c|c}\hline
$l_{\infty}$-error &$\Delta t=0.001$&$\Delta t=0.002$&$\Delta t=0.0025$&$\Delta t=0.004$& $\Delta t=0.005$&rate \\\hline

$t=10$    & 2.44e-08 &  9.76e-08 & 1.52e-07  & 3.90e-07 & 6.10e-07 &2.00\\\hline
$t=15$    & 7.90e-08 &  3.16e-07 & 4.93e-07  & 1.26e-06& 1.97e-06&2.00\\\hline
$t=40$     & 8.13e-07 &  3.25e-06 & 5.08e-06  & 1.30e-05 & 2.03e-05 &2.00 \\\hline
$t=100$    & 3.90e-07 &  1.56e-06 & 2.44e-06  & 6.24e-06 & 9.75e-06 &2.00\\\hline
\end{tabular}
\end{center}
\end{table}

\subsection{Seismic source}\label{Seis}
\begin{figure}\centering
\includegraphics[width=10cm]{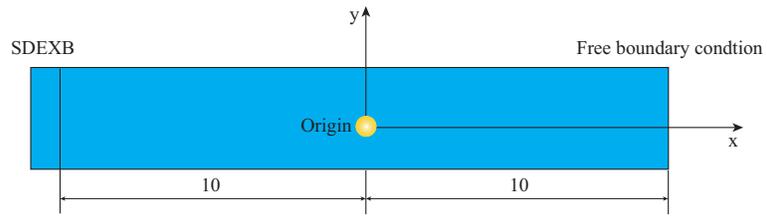}
\caption{A segment of an infinite bar with a Ricker pulse.}\label{or}
\end{figure}

We now study a 1D semi-infinite domain perturbed by a pulse. Following \cite{Linjuan}, the computational domain is chosen by $x\in [-10,10]$ just as Figure \ref{or} and the micromodulus function is chosen as (\ref{modu}). We use semi-discrete exact boundary at the left end, and the right end is a free boundary. The pulse is set at origin, can be written as
\begin{eqnarray}\label{seismic}
&&u_{0}(t)=(1-2\pi^2f_{p}^2(t-t_{D})^2)e^{-\pi^2f_{p}^2(t-t_{D})^2},
\end{eqnarray}
where $f_{p}$ is the peak frequency and $t_{D}$ is the delay time. The dimensionless parameters are $f_{p}=0.2$, $t_{D}=5$. Other settings are the same as section \ref{bar}. We remark that
the pulse works only for $t\in [0, 5]$. When $t>5$, the system will return to free vibration.

The numerical results and reference solutions are plotted in Figure \ref{urml}. The displacements of the origin, the left end, and the right end
are depicted. One can see a good agreement.

At beginning, the pulse at the origin will propagates
to the two sides. At $t=5$ pulse will reach its peak as shown in Figure \ref{urml}(a).

Then as shown in
Figure \ref{urml}(b) the pulse reaches the right boundary at
about $t=15$. Because of the free boundary condition, the amplitude of the right end is almost doubled. In the same time $15$ the pulse which is the first peak as shown in Figure \ref{urml}(c) passes through the left end.

After that the reflection wave at right end will go back and never comes again, therefore the peak in Figure \ref{urml}(b) is the only peak of right end. Then the reflection wave generated by the right free boundary passes through the origin, the reflection wave is the second hump shown in Figure \ref{urml}(a).

Then the reflection wave passes through the left boundary, the reflection wave is the second hump shown in Figure \ref{urml}(c).
In Figure \ref{urml}(b) there is only one hump, which means there is no visible reflection wave from the left boundary. It clearly shows the effectiveness of exact boundary conditions from the comparison with reference solutions in Figure \ref{urml}.

\begin{figure}\centering \label{urml}
\includegraphics[width=10cm]{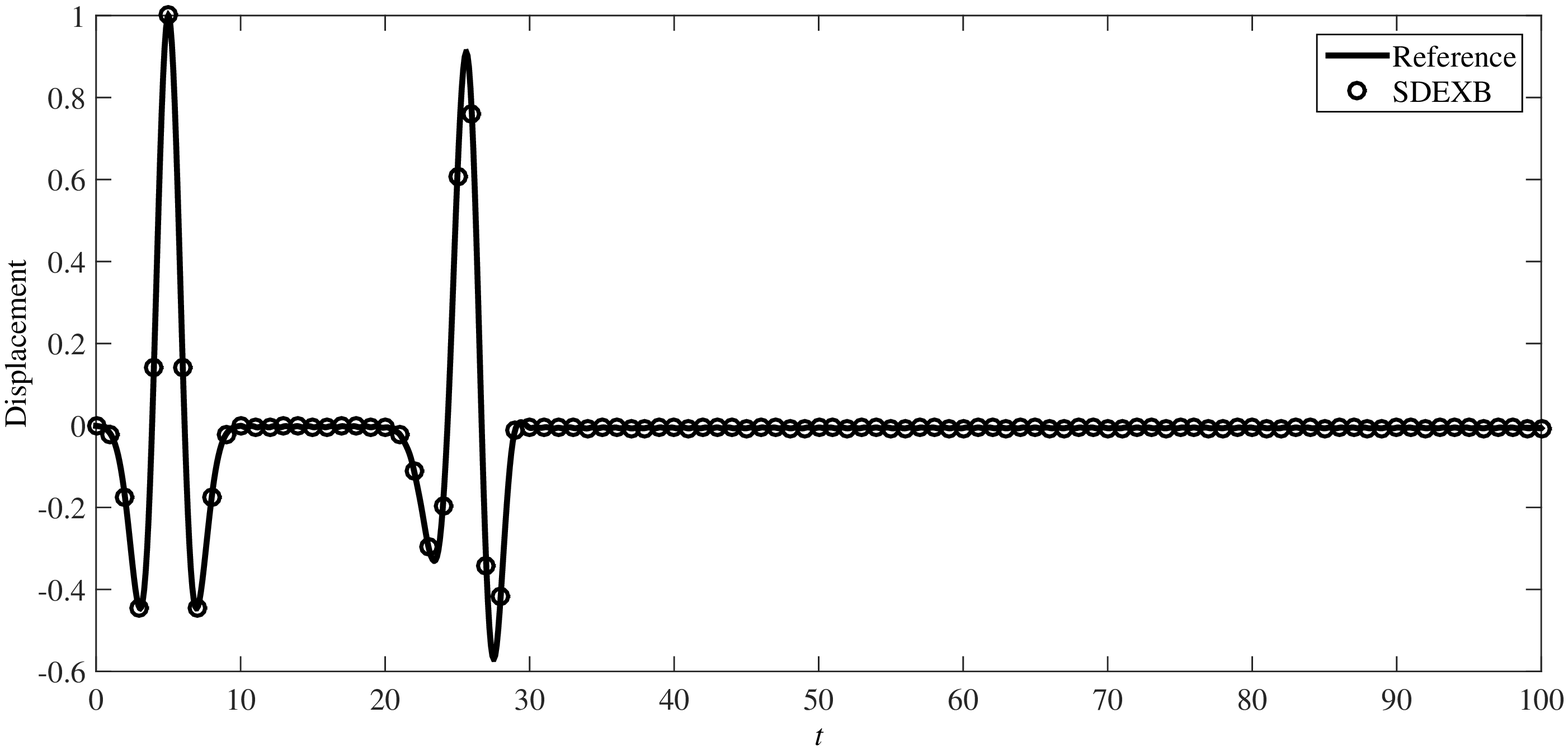}
\includegraphics[width=10cm]{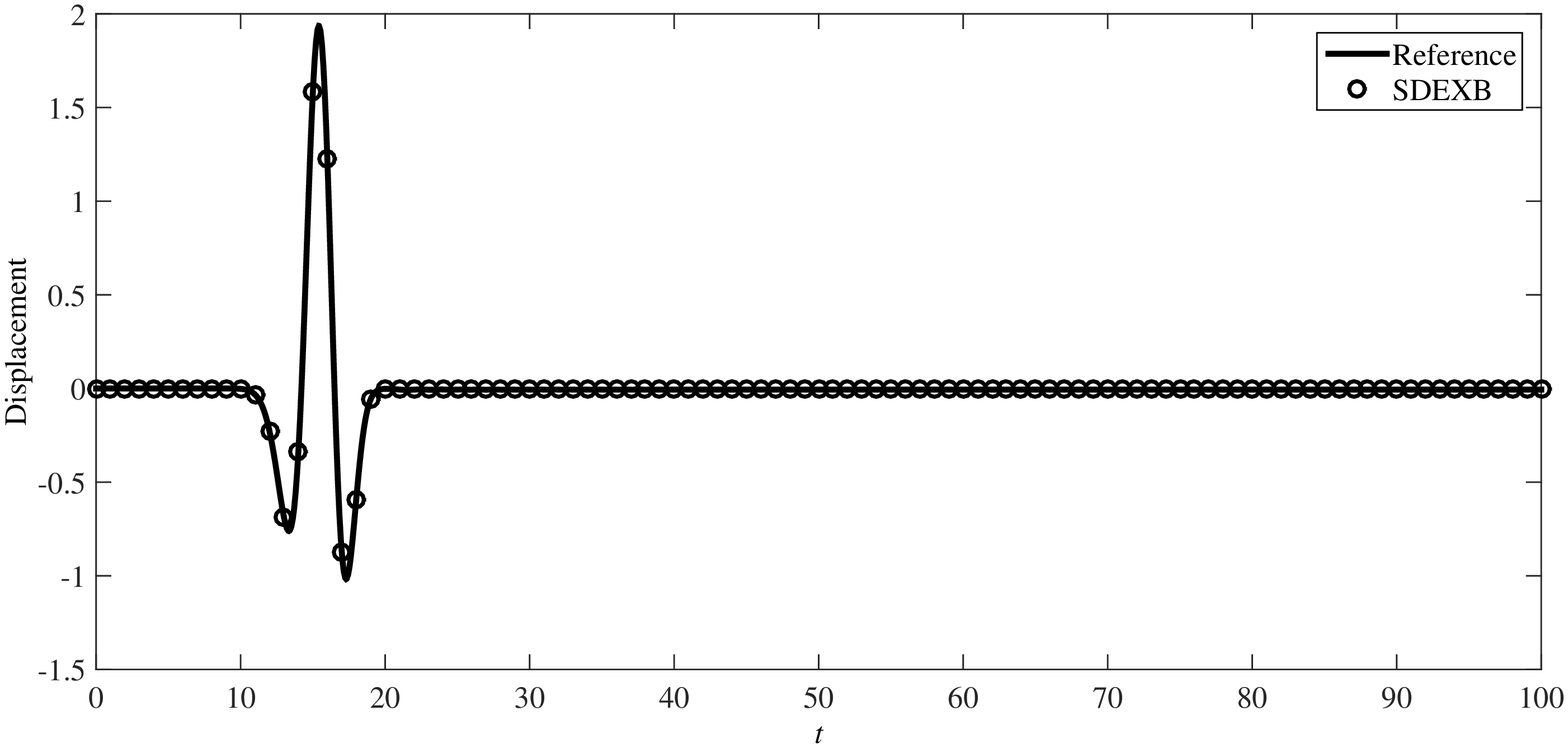}
\includegraphics[width=10cm]{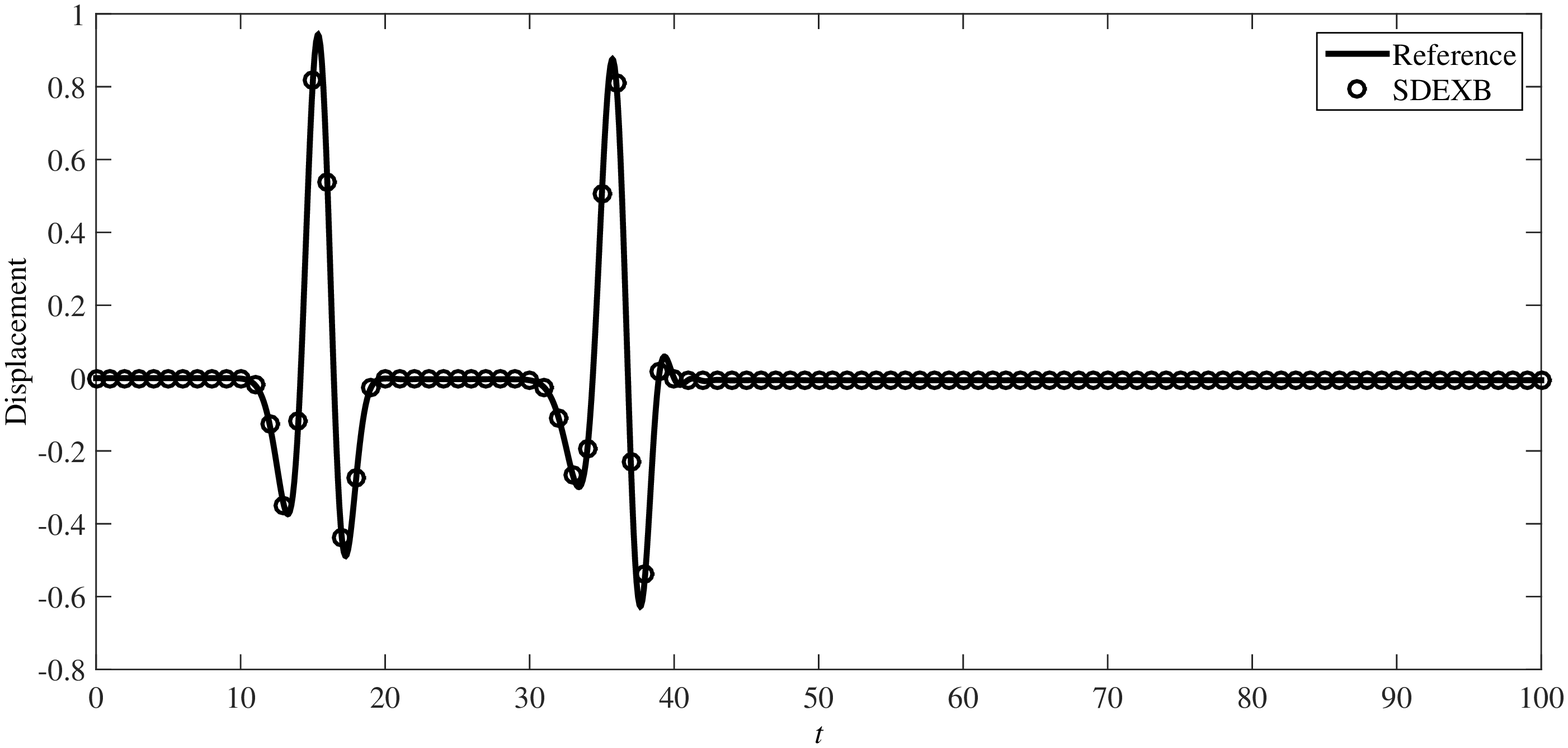}
\caption{Displacements of three typical points: (a) the origin; (b) the right end; (b) the left end.}\label{urml}
\end{figure}
\begin{figure}\centering
\includegraphics[width=8cm]{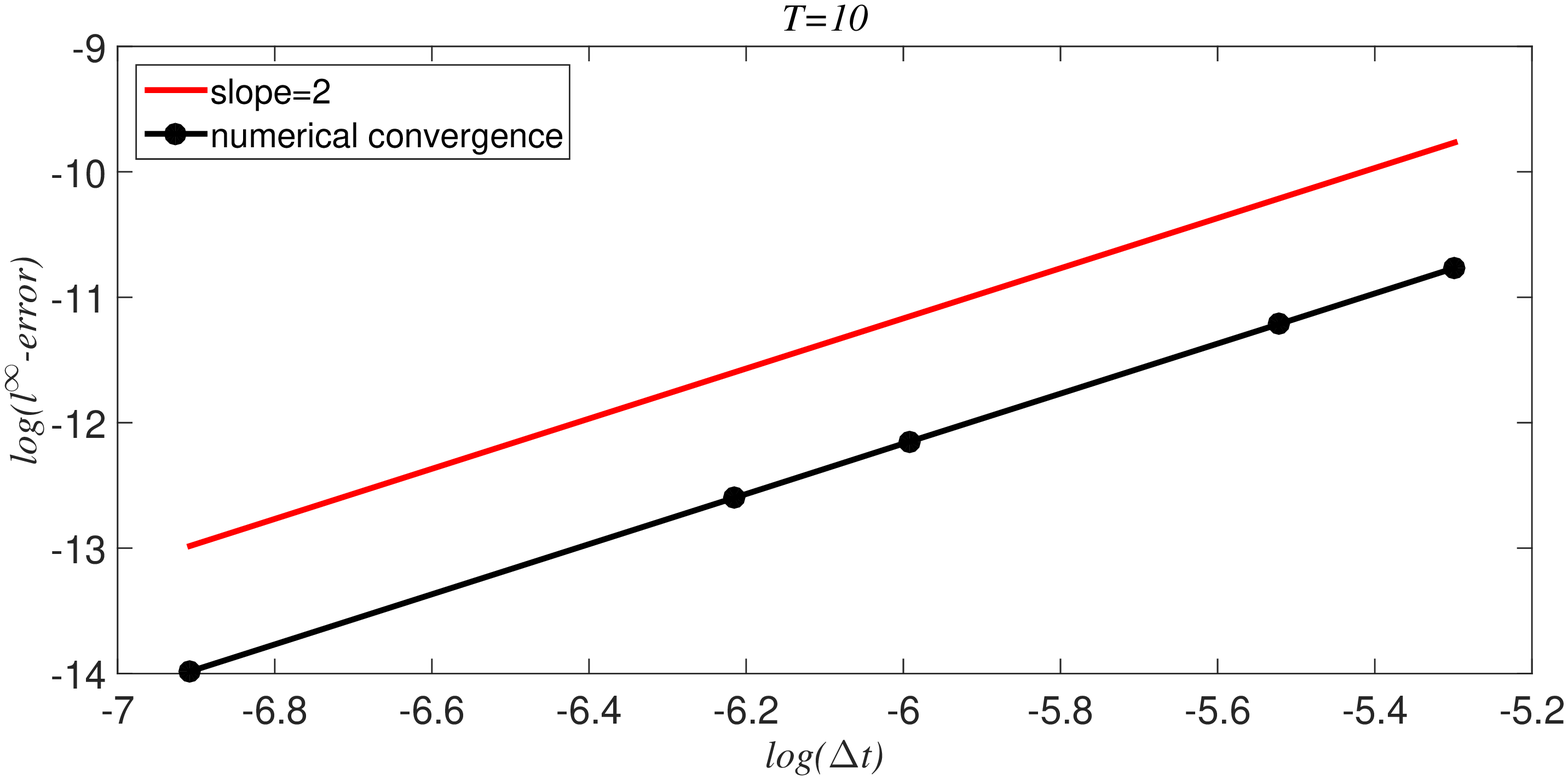}
\includegraphics[width=8cm]{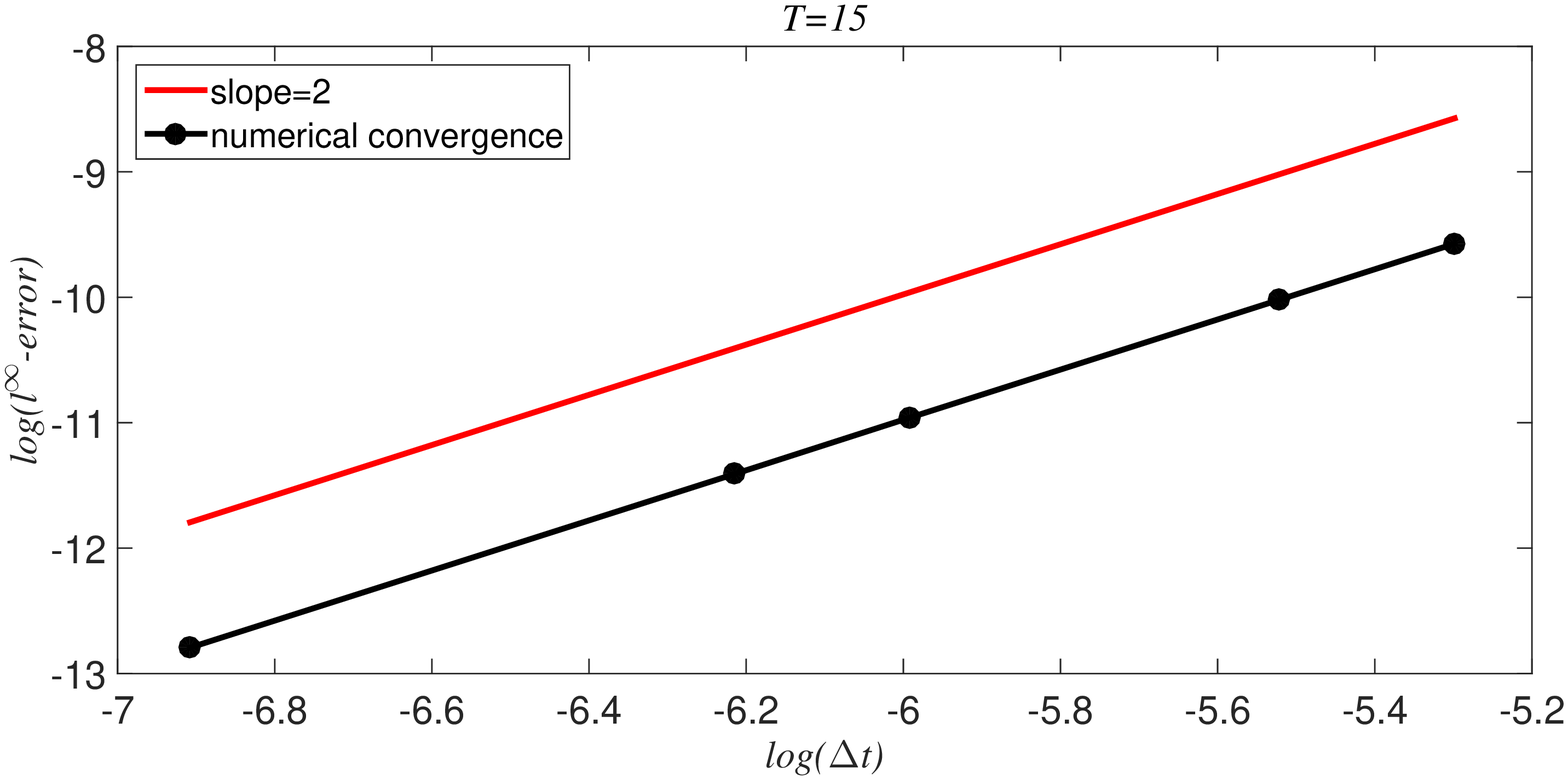}
\includegraphics[width=8cm]{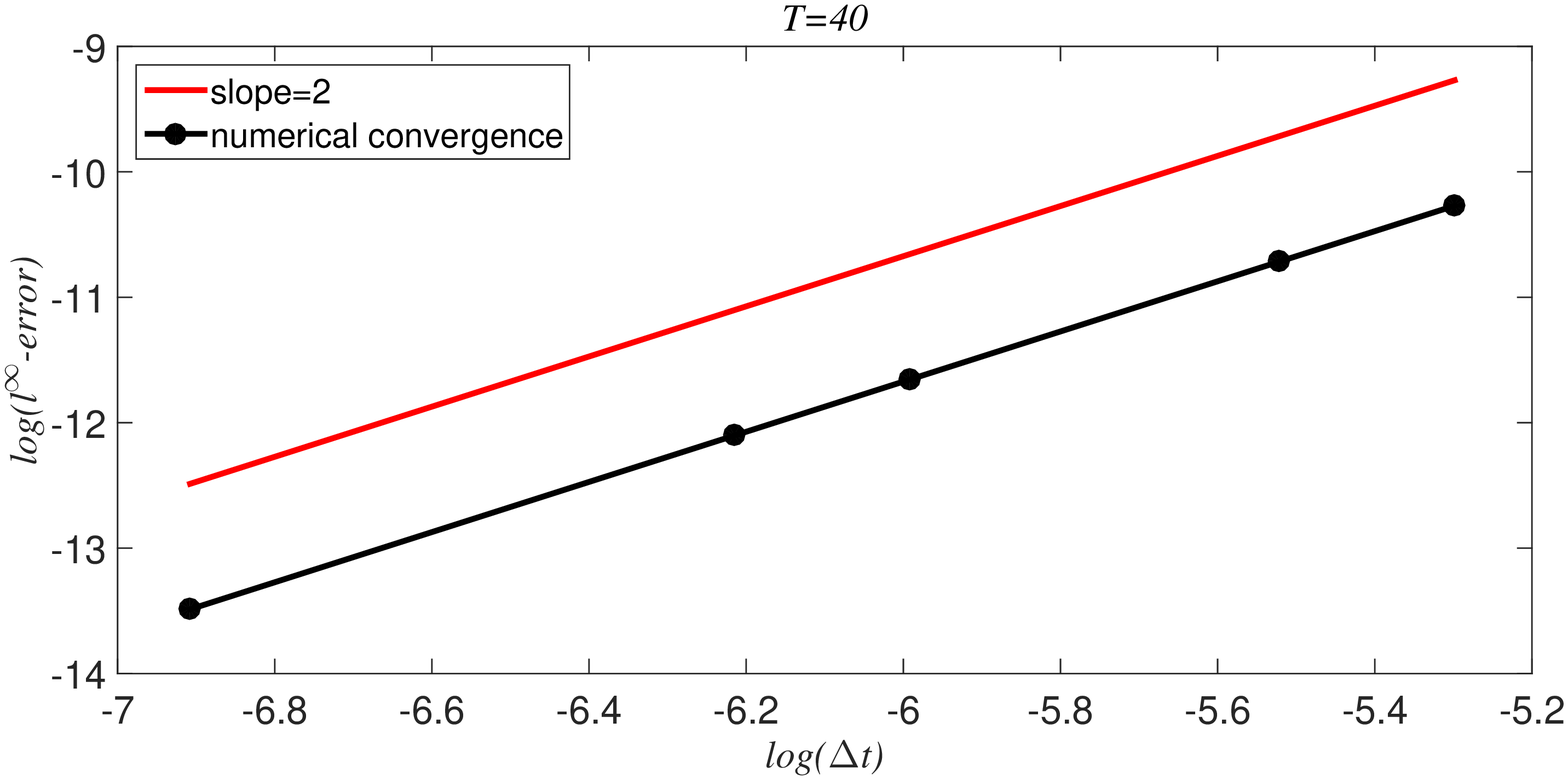}
\includegraphics[width=8cm]{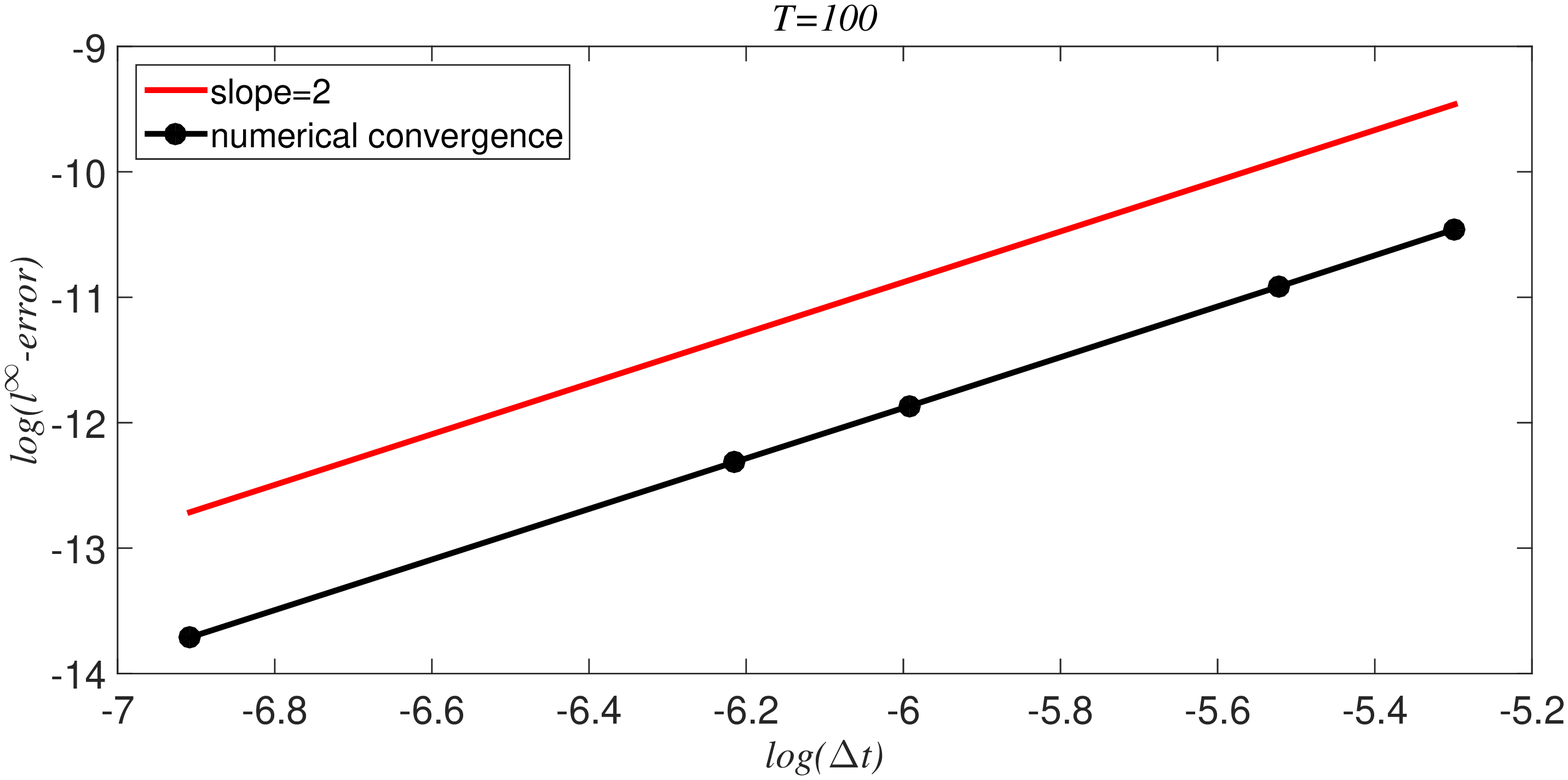}
\caption{Convergence rate of displacement at different time.}\label{ueprob3}
\end{figure}

The $l_{\infty}$-error of left end's displacements is used to measure the performance. The $l^{\infty}$ error is listed in Tables \ref{tabeuler3}. The convergence rates at different times are around second order. The convergence rates of displacement are depicted in Fig \ref{ueprob3}.

\begin{table}[h]
\begin{center}
\caption { $l_{\infty}$-error and convergence rate of displacement for different time.}\label{tabeuler3}
\begin{tabular}{|c|c|c|c|c|c|c}\hline
$l_{\infty}$-error &$\Delta t=0.001$&$\Delta t=0.002$&$\Delta t=0.0025$&$\Delta t=0.004$& $\Delta t=0.005$&rate \\\hline

$t=10$    & 8.46e-07 &  3.38e-06 & 5.29e-06  & 1.35e-05 & 2.11e-05 &2.00\\\hline
$t=15$    & 2.78e-06 &  1.11e-05 & 1.74e-05  & 4.45e-05& 6.96e-05&2.00\\\hline
$t=40$     & 1.38e-06 &  5.55e-06 & 8.67e-06  & 2.21e-05 & 3.47e-05 &2.00 \\\hline
$t=100$    & 1.11e-06 &  4.49e-06 & 7.03e-06  & 1.82e-05 & 2.87e-05 &2.02\\\hline
\end{tabular}
\end{center}
\end{table}

\section{Conclusion and future work}\label{sec:conclusion}

 Semi-discrete exact artificial boundary conditions (SDEXABCs) for peridynamics are proposed. The kernel functions are obtained from integral system by introducing auxiliary functions $g_{n}(t)$. Fortunately the $g_{n}(t)$ can be computed from ordinary differential system by RK4 through a recursive
relationship. The kernel functions are combined with the construct  of SDEXABCs. We use the boundary conditions and the Verlet algorithm  to solve the peridynamics.  Second order accuracy is confirmed.

The algorithm for computing kernel functions avoids the numerical Fourier transform. Generally the numerical Fourier transform for solving kernel functions brings large numerical error in long-time simulations. The algorithm only depends on the ordinary differential system and integral system that can be handled by high-precision algorithm.

The numerical algorithm may be extended into higher dimensional
case. We further remark that this approach may be extended for more general nonlocal PDE, as well as for more crystal lattice systems in computational mechanics.

Though the exact boundary conditions of semi-discrete peridynamics are obtained; some questions regarding theoretical and numerical aspects of these boundary conditions remain
to be answered. Such as the extension of high dimensional cases which will be reported in our forthcoming paper, and the rigorous proof of the numerical convergence rate of the CN scheme. How to reduce the huge amount of computation is still an open question, the fast algorithms of the boundary conditions are also
needed to be explored. Machine learning may be used to deal with the problems, one would refer \cite{YP1,YP2,YP3}. We intend to address the question in future.

\section*{Acknowledgments}

This research is partially supported by NSFC under grant Nos. 11502028. Yibo Yang and Paris Perdikaris are supported by the Advanced Scientific Computing Research program (grant DE-SC0019116) and the Defense Advanced Research Projects Agency under the Physics of Artificial Intelligence program (grant HR00111890034).  The authors thank for Dr Linjuan Wang for the useful helps and suggestions.

\section*{References}

%% not want to use model1c-num-names.bst.


%\bibliographystyle{plain}
%\bibliography{HighOrderCaputo}
\bibliographystyle{unsrt}
%\bibliography{HighOrderCaputo}

\begin{thebibliography}{}
\bibitem{Sill}
S Silling. Reformulation of elasticity theory for discontinuities and long-range forces, Journal of the Mechanics
and Physics of Solids. 48 (2000) 175-209.
\bibitem{Gerstle}
W Gerstle, N Sau, S Silling. Peridynamic modeling of concrete structures, Nuclear Engineering and Design. 237 (2007) 1250-1258.
\bibitem{Xu}
J Xu, A Askari, O Weckner, S Silling. Peridynamic analysis of impact damage in composite laminates, Journal
of Aerospace Engineering. 21 (2008) 187-194.
\bibitem{Kilic}
B Kilic, A Agwai, E Madenci. Peridynamic theory for progressive damage prediction in center-cracked composite
laminates, Composite Structures. 90 (2009) 141-151.
\bibitem{Oterkus}
E Oterkus, E Madenci. Peridynamic analysis of fiber-reinforced composite materials, Journal of Mechanics of
Materials and Structures. 70 (2012) 45-84.
\bibitem{Oterkus2}
E Oterkus, E Madenci. Peridynamic theory for damage initiation and growth in composite laminate, Advances in
Fracture and Damage Mechanics. 488 (2012) 355-358.
\bibitem{Kelle}
J Kelle, D Givoli. Exact non-reflecting boundary-conditions, Journal of Computational Physics. 82 (1989) 172-192.
\bibitem{Berenger}
J Berenger. A perfectly matched layer for the absorption of electromagnetic-waves, Journal of Computational
Physics. 114 (1994) 185-200.
\bibitem{Weckner}
O Weckner, R Abeyaratne. The effect of long-range forces on the dynamics of a bar, Journal of the Mechanics and
Physics of Solids. 53 (2005) 705-728.
\bibitem{Silling2}
S Silling, M Zimmermann, R Abeyaratne. Deformation of a peridynamic bar, Journal of Elasticity. 73 (2003) 173-190.
\bibitem{Weckner2}
O Weckner, E Emmrich. Numerical simulation of the dynamics of a nonlocal, inhomogeneous, infinite bar, Journal
of Computational and Applied Mechanics. 6 (2005) 311-319.
\bibitem{Emmrich}
E Emmrich, O Weckner. Analysis and numerical approximation of an integro-differential equation modeling non-local
effects in linear elasticity, Mathematics and Mechanics of Solids. 12 (2007) 363-384.
\bibitem{Mikata}
Y Mikata. Analytical solutions of peristatic and peridynamic problems for a 1D infinite rod, International Journal of
Solids and Structures. 49 (2012) 2887-2897.
\bibitem{Liao}
Z Liao, H Wong. A transmitting boundary for the numerical simulation of elastic wave propagation, International
Journal of Soil Dynamics and Earthquake Engineering. 3 (1984) 174-183.
\bibitem{Fevens}
T Fevens, H Jiang. Absorbing Boundary Conditions for the Schr\"odinger Equation, Siam Journal on Scientific Computing. 21 (1999) 255-282.
\bibitem{Xavier1}
X Antoine, C Besse, P Klein. Absorbing boundary conditions for the one-dimensional Schr\"odinger equation with an exterior repulsive potential, Journal of Computational Physics. 228 (2009) 312-335.
\bibitem{Xavier2}
X Antoine, C Besse. Unconditionally stable discretization schemes of non-reflecting boundary conditions for the one-dimensional Schr\"odinger equation, Journal of Computational Physics. 188 (2003) 157-175.
\bibitem{Anton}
A Arnold, M Ehrhardt, I Sofronov. Discrete transparent boundary conditions for the Schr\"odinger equation: fast calculation, approximation, and stability, Communications in Mathematics Science. 1 (2004) 501-556.
\bibitem{Greengard}
L Greengard, P Lin. On the numerical solution of the heat equation in
unbounded domains (Part I), Tech. Note 98002, Courant Mathematics and
Computing Laboratory, New York University. 1998.
\bibitem{Wu}
X Wu, Z Sun. Convergence of difference scheme for heat equation in unbounded domains using artificial boundary conditions, Applied Numerical Mathematics. 50 (2004) 261-277.
\bibitem{Baskakov}
V Baskakov, A Popov. Implementation of transparent boundaries for numerical solution of the Schr\"odinger equation, Wave Motion. 14 (1991) 123-128.
\bibitem{Han1}
H Han, Z Huang. Exact and approximating boundary conditions for the parabolic problems on unbounded domains, Computers and Mathematics with Applications. 44 (2002) 655-666.
\bibitem{Han2}
H Han, Z Huang. Exact artificial boundary conditions for the Schr\"odinger equation in $R^2$, Communications in Mathematical Sciences. 2 (2004)  79-94.
\bibitem{Anton2}
A Arnold, M Ehrhardt, M Schulte, I Sofronov. Discrete transparent boundary conditions for the Schr\"odinger equation on circular domains, Communications in Mathematical Sciences. 10 (2012) 889-916.
\bibitem{Li2}
H Li, X Wu, J Zhang. Local artificial boundary conditions for Schr\"odinger and heat equations by using high-order azimuth derivatives on circular artificial boundary, Computer Physics Communications. 185 (2014) 1606-1615.
\bibitem{Besse}
X Antoine, C Besse, V Mouysset. Numerical schemes for the simulation of the two-dimensional Schr\"odinger equation using non-reflecting boundary conditions, Mathematics of Computation. 73 (2004) 1779-1799.
\bibitem{Besse2}
X Antoine, C Besse, P Klein. Absorbing Boundary Conditions for the Two-Dimensional Schr\"odinger Equation with an Exterior Potential. Part II: Discretization and Numerical Results, Numerische Mathematik. 125 (2013) 191-223.
\bibitem{Besse3}
X Antoine, C Besse, P Klein. Absorbing Boundary Conditions for General Nonlinear Schr\"odinger Equations, Siam Journal on Scientific Computing. 33 (2011) 1008-1033.
\bibitem{Kadowaki}
H Kadowaki, W Liu. A multiscale approach for the micropolar continuum model, CMES-Computer Modeling in Engineering and Sciences. 7 (2005) 269-282.
\bibitem{Pangs}
G Pang, Y Yang, A Xavier, S Tang. Stability and convergence analysis of artificial boundary conditions for the Schr{\"o}dinger equation on a rectangular domain. Submitted.
\bibitem{Besse4}
X Antoine, A Arnold, C Besse. A Review of Transparent and Artificial Boundary Conditions Techniques for Linear and Nonlinear Schr\"odinger Equations, Communications in Computational Physics. 4 (2008) 729-796.
\bibitem{Adelman}
S Adelman, J Doll. Generalized Langevin equation approach for atom/solid-surface scattering-collinear atom/harmonic chain model, Journal of Chemical Physics. 61 (1974) 4242-4245.
\bibitem{Cai}
W Cai, D Koning, V Bulatov, S Yip. Minimizing boundary reflections in coupled-domain simulations, Physical
Review Letters. 2000 85(2000) 3213-3216.
\bibitem{Kadowaki}
H Kadowaki, W Liu. A multiscale approach for the micropolar continuum model, CMES-Computer Modeling in
Engineering and Sciences 2005. 7 (2005) 269-282.
\bibitem{Karpov}
E Karpov, G Wagner, W Liu. A Green's function approach to deriving non-reflecting boundary conditions in
molecular dynamics simulations, International Journal for Numerical Methods in Engineering. 62 (2005) 250-1262.
\bibitem{Park}
H Park, E Karpov, P Klein, W Liu. Three-dimensional bridging scale analysis of dynamic fracture, Journal of
Computational Physics. 207 (2005) 588-609.
\bibitem{Park1}
H Park, E Karpov, W Liu. Non-reflecting boundary conditions for atomistic, continuum and coupled atomistic/continuum simulations, International Journal for Numerical Methods in Engineering. 64 (2005) 237-259.
\bibitem{Engquist}
B Engquist, A Majda. Radiation boundary conditions for acoustic and elastic calculations, Communications on Pure and Applied Mathematics. 32 (1979) 313-357.
\bibitem{E}
W E, Z Huang. Matching conditions in atomistic-continuum modeling of materials, Physical Review Letters. 87 (2001) Art135501(1-4).
\bibitem{Li}
X Li, W E. Variational boundary conditions for molecular dynamic simulation of solids at low temperature, Communications in Computational Physics. 1 (2006) 135-175.
\bibitem{Jone}
R Jones, C Kimmer. Efficient non-reflecting boundary condition constructed via optimization of damped layers, Physical Review B. 81 (2010) 094301.
\bibitem{Tang2}
S Tang. A finite difference approach with velocity interfacial conditions for multiscale computations of crystalline solids, Journal of Computational Physics. 227 (2008) 4038-4062.
\bibitem{Tang3}
X Wang, S Tang. Matching boundary conditions for diatomic chains, Computational Mechanics. 46 (2010) 813-826.
\bibitem{Tang4}
X Wang, S Tang. Matching boundary conditions for lattice dynamics, International Journal
for Numerical Methods in Engineering. 93 (2013) 1255-1285.
\bibitem{Tang5}
G Pang, S Tang. Approximate linear relations for Bessel functions, Communications in Mathematical Sciences. 15 (2017) 1967-1986.
\bibitem{Tang6}
G Pang, L Bian, S Tang. ALmost EXact boundary condition for one-dimensional Schr\"odinger
equation, Physical Review E. 86 (2012) 066709.
\bibitem{Linjuan}
L Wang, Y Chen, J Xu, Jianxiang Wang. Transmitting boundary conditions for 1D peridynamics, International Journal for Numerical Methods in Engineering. 110 (2017) 379-400.
\bibitem{Wang}
X Wang, S Tang. Matching boundary conditions for lattice dynamics, International Journal for Numerical Methods
in Engineering. 93 (2013) 1255-1285.
\bibitem{Silling3}
S Silling, A Askari. A meshfree method based on the peridynamic model of solid mechanics, Computers and
Structures. 83 (2005) 1526-1535.
\bibitem{Zhang}
C Zheng, J Hu, Q Du, J Zhang. Numerical solution of the nonlocal diffusion equation on the real line, Siam Journal on Scientific Computing. 39 (2017) 1951-1968.
\bibitem{Zhang2}
J Wang, J Zhang, C Zheng. Stability and error analysis for a second-order approximation of the 1D nonlocal Schr\"odinger equation under DtN-type boundary conditions, under review.
\bibitem{YP4}
S Tang, Y Ying, Y Lian, W Liu. Differential operator multiplication method for fractional differential equations, Computational Mechanics. 58(6) (2016) 879-888.
\bibitem{Pang}
S Ji, Y Yang, G Pang, X Antoine. Accurate artificial boundary conditions for the semi-discretized linear
Schr\"odinger and heat equations on rectangular domains, Computer Physics Communications. 222 (2018) 84-93.
\bibitem{Pang2}
G Pang, S Ji, Y Yang, S Tang. Eliminating corner effects in square lattice simulation, Computaional Mechanics. 62 (2018) 111-122.
\bibitem{Silling4}
S Silling. Origin and effect of nonlocality in a composite, Journal of Mechanics of Materials and Structures. 9 (2014) 245-257.
\bibitem{Oterkus3}
E Madenci, E Oterkus. Peridynamic Theory and its Applications, Springer: New York, 2014.
\bibitem{Wangjihong}
J Wang, J Zhang, C Zheng. Stability and error analysis for a second-order approximate of the 1D nonlocal Schr\"odinger equation under DtN-type boundary conditions, Submitted.
\bibitem{Duwave}
Q Du, H Han, J Zhang, C Zheng. Numerical Solution of a two-dimensional nonlocal wave equation on unbounded domains, Siam Journal on Scientific Computing. 40 (2018) 1430-1445.
\bibitem{YP1}
G Kissas, Y Yang, E Hwuang
P Perdikaris. Machine learning in cardiovascular flows modeling: Predicting arterial blood pressure from non-invasive 4D flow MRI data using physics-informed neural networks, Submitted.
\bibitem{YP2}
Y Yang, P Perdikaris. Adversarial uncertainty quantification in physics-informed neural networks, Journal of Computational Physics. 394 (2019) 136-152.
\bibitem{YP3}
Y Yang, P Perdikaris. Physics-informed deep generative models, Nips. 2018.


%\bibitem{E2}
%Weinan E, Huang ZY. A dynamic atomistic-continuum method for the simulation of crystalline materials. Journal of
%Computational Physics. 182 (2002) 234-261.
%
%\bibitem{Bobaru}
%Bobaru F, Yang M, Alves LF, Silling SA, Askari E, Xu JF. Convergence, adaptive refinement, and scaling in 1D
%peridynamics. International Journal for Numerical Methods in Engineering. 77 (2009) 852??C877.
%\bibitem{Ha}
%Ha YD, Bobaru F. Studies of dynamic crack propagation and crack branching with peridynamics. International
%Journal of Fracutre. 162 (2010) 229??C244.
%
%
%\bibitem{LiX}
%X Li, W E. Variational boundary conditions for molecular dynamics simulation of solids at low
%temperature. Communications in Computational Physics. 1 (2006) 135-175.
%
%
%
%
%\bibitem{Ha2}
%Ha YD, Bobaru F. Characteristics of dynamic brittle fracture captured with peridynamics. Engineering Fracture
%Mechanics. 78 (2011) 1156??C1168.
%\bibitem{Chen}
%Chen Z, Bobaru F, Zhang G. Selecting the kernel in a peridynamic formulation: a study for transient heat diffusion.
%Computer Physics Communications. 197 (2015) 51??C60.
%\bibitem{Bobaru2}
%Bobaru F, Zhang G. Why do cracks branch? A peridynamic investigation of dynamic brittle fracture. International
%Journal of Fracutre. 196 (2015) 59??C98.
%
%\bibitem{Silling5}
%Silling SA. Linearized theory of peridynamic states. Journal of Elasticity. 93 (2010) 1255??C1285.
%\bibitem{Madenci}
%Madenci E, Oterkus E. Peridynamic Theory and its Applications. Springer: New York, 2014.
%\bibitem{Liao2}
%Liao ZP. Introduction to Wave Motion Theories in Engineering (In Chinese). Science Press: Beijing, 2002.


\end{thebibliography}

{}



% \end{thebibliography}

\end{document}